\documentclass[twocolumn,aps,prc,superscriptaddress,showpacs,floatfix]{revtex4}
\usepackage{amssymb}
\usepackage{amsmath}
\usepackage{graphicx}
\usepackage[normalem]{ulem}
\usepackage[dvips]{color}

\renewcommand\sout{\bgroup \color{red} \ULdepth=-.5ex \ULset}

\begin{document}

\title{Isospin-dependent properties of asymmetric nuclear matter in
relativistic mean-field models}
\author{Lie-Wen Chen}
\affiliation{Institute of Theoretical Physics, Shanghai Jiao Tong
University, Shanghai 200240, China} \affiliation{Center of
Theoretical Nuclear Physics, National Laboratory of Heavy Ion
Accelerator, Lanzhou 730000, China}
\author{Che Ming Ko}
\affiliation{Cyclotron Institute and Physics Department, Texas A\&M
University, College Station, Texas 77843-3366, USA}
\author{Bao-An Li}
\affiliation{Department of Physics, Texas A\&M University-Commerce,
Commerce, Texas 75429-3011, USA}
\date{\today }

\begin{abstract}
Using various relativistic mean-field models, including the
nonlinear ones with meson field self-interactions, those with
density-dependent meson-nucleon couplings, and the point-coupling
models without meson fields, we have studied the isospin-dependent
bulk and single-particle properties of asymmetric nuclear matter.
In particular, we have determined the density dependence of
nuclear symmetry energy from these different relativistic
mean-field models and compare the results with the constraints
recently extracted from analyses of experimental data on isospin
diffusion and isotopic scaling in intermediate-energy heavy ion
collisions as well as from measured isotopic dependence of the
giant monopole resonances in even-A Sn isotopes. Among the $23$
parameter sets in the relativistic mean-filed model that are
commonly used for nuclear structure studies, only a few are found
to give symmetry energies that are consistent with the empirical
constraints. We have also studied the nuclear symmetry potential
and the isospin-splitting of the nucleon effective mass in isospin
asymmetric nuclear matter. We find that both the momentum
dependence of the nuclear symmetry potential at fixed baryon
density and the isospin-splitting of the nucleon effective mass in
neutron-rich nuclear matter depend not only on the nuclear
interactions but also on the definition of the nucleon optical
potential.
\end{abstract}

\maketitle

\section{Introduction}

Besides the many existing radioactive beam facilities and their
upgrades, many more are being constructed or under planning,
including the Cooling Storage Ring (CSR) facility at HIRFL in
China \cite{CSR}, the Radioactive Ion Beam (RIB) Factory at RIKEN
in Japan \cite{Yan07}, the FAIR/GSI in Germany \cite{FAIR},
SPIRAL2/GANIL in France \cite{SPIRAL2}, and the Facility for Rare
Isotope Beams (FRIB) in the USA \cite{RIA}. These new facilities
offer the possibility to study the properties of nuclear matter or
nuclei under the extreme condition of large isospin asymmetry. As
a result, the study of the isospin degree of freedom in nuclear
physics has recently attracted much attention. The ultimate goal
of such study is to extract information on the isospin dependence
of in-medium nuclear effective interactions as well as the
equation of state (EOS) of isospin asymmetric nuclear matter,
particularly its isospin-dependent term or the density dependence
of the nuclear symmetry energy. This knowledge, especially the
latter, is important for understanding not only the structure of
radioactive nuclei, the reaction dynamics induced by rare
isotopes, and the liquid-gas phase transition in asymmetric
nuclear matter, but also many critical issues in astrophysics
\cite{LiBA98,LiBA01b,Dan02a,Lat00,Lat01,Lat04,Bar05,Ste05a}.
Unfortunately, the density dependence of the nuclear symmetry
energy, especially its behavior at high densities, is largely
unknown and is regarded as the most uncertain among all the
properties of isospin asymmetric nuclear matter. Although the
nuclear symmetry energy at normal nuclear matter density $\rho
_{0}\approx 0.16$ fm$^{-3}$ is known to be around $30$ MeV from
the empirical liquid-drop mass formula \cite{Mey66,Pom03}, its
values at other densities are poorly known \cite{LiBA98,LiBA01b}.
Various microscopic and phenomenological models, such as the
relativistic Dirac-Brueckner-Hartree-Fock (DBHF)
\cite{Ulr97,Fuc04,Ma04,Sam05a,Fuc05,Fuc05b,Ron06} and the
non-relativistic Brueckner-Hartree-Fock (BHF) \cite{Bom91,Zuo05}
approach, the relativistic mean-field (RMF) model based on
nucleon-meson interactions \cite{Bar05}, and the non-relativistic
mean-field model based on Skyrme-like interactions
\cite{Das03,LiBA04a,LiBA04c,Che04,Riz04,Beh05,Riz05}, have been
used to study the isospin-dependent properties of asymmetric
nuclear matter, such as the nuclear symmetry energy, the nuclear
symmetry potential, the isospin-splitting of nucleon effective
mass, etc., but the predicted results vary widely. In fact, even
the sign of the symmetry energy above $3\rho _{0}$ is uncertain
\cite{Bom01}. The theoretical uncertainties are mainly due to the
lack of knowledge about the isospin dependence of in-medium
nuclear effective interactions and the limitations in the
techniques for solving the nuclear many-body problem. As to the
incompressibility of asymmetric nuclear matter, it is essentially
undetermined \cite{Shl93}, even after about 30 years of studies.
For comparison, the incompressibility of symmetric nuclear matter
at its saturation density $\rho_0$ has been determined to be
$231\pm 5$ MeV from the nuclear GMR \cite{You99} and the EOS at
densities of $2\rho _{0}<\rho <5\rho _{0}$ has also been
constrained by measurements of collective flows in nucleus-nucleus
collisions \cite{Dan02a}.

As a phenomenological approach, the RMF model has achieved great
success during the last decade in describing many nuclear
phenomena \cite{Wal74,Ser86,Rei89,Rin96,Ser97,Ben03,Fur04,Men06}.
For example, it provides a novel saturation mechanism for the
nuclear matter, an explanation of the strong spin-orbit
interaction in finite nuclei, a natural energy dependence of the
nucleon optical potential, etc. The RMF approach is generally
based on effective interaction Lagrangians that involve nucleon
and meson fields. In this approach, a number of parameters are
adjusted to fit the properties of many nuclei. As such, these
models usually give excellent descriptions of nuclear properties
around or below the saturation density.

Since the original Lagrangian proposed by Walecka more than $30$
years ago \cite{Wal74}, there have been a lot of different
treatments, extensions, and applications of the RMF model. The
three main versions are the nonlinear models
\cite{Ser86,Rei89,Rin96,Ser97}, models with density-dependent
meson-nucleon couplings \cite{Len95,Fuc95,She97,Typ99,Hof01}, and
point-coupling models without mesons
\cite{Nik92,Bur02,Mad04,Bur04,Fin04,Fin06}. For each version of
the RMF model, there are also many different parameter sets with
their values fitted to the binding energies and charge radii of a
large number of nuclei in the periodic table. Including isovector
mesons in the effective interaction Lagrangians further allows the
RMF model to describe successfully the properties of nuclei far
from the $\beta $-stability line. With recent developments in
constraining the isospin-dependent properties of asymmetric
nuclear matter, especially the density dependence of the nuclear
symmetry energy, it is of great interest to see to what extend the
results from different versions of the RMF model are consistent
with these constrains.

In the present work, based on commonly used $23$ different
parameter sets in three different versions of the RMF model, we
carry out a systematic study of the isospin-dependent bulk and
single-particle properties of asymmetric nuclear matter. In
particular, we study the density dependence of the nuclear
symmetry energy from these RMF models and compare the results with
the constraints recently extracted from analyses of the isospin
diffusion data from heavy-ion collisions based on the isospin and
momentum-dependent IBUU04 transport model with in-medium
nucleon-nucleon (NN) cross sections \cite{Tsa04,Che05a,LiBA05c},
the isoscaling analyses of isotope ratios in intermediate energy
heavy ion collisions \cite{She07}, and measured isotopic
dependence of the giant monopole resonances (GMR) in even-A Sn
isotopes \cite{Gar07}. Among these $23$ commonly used interactions
in nuclear structure studies, only a few are found to give
symmetry energies that are consistent with the empirically
extracted one. Furthermore, we study the nuclear symmetry
potential and the isospin-splitting of the nucleon effective mass
in isospin asymmetric nuclear matter. Our results indicate that
the nuclear symmetry potential at fixed baryon density may
increase or decrease with increasing nucleon momentum depending on
the definition of the nucleon optical potential and the
interactions used. This dependence is also seen in the
isospin-splitting of the nucleon effective mass in neutron-rich
nuclear matter. In addition, the isospin-splitting of the nucleon
scalar density in neutron-rich nuclear matter is also studied.

The paper is organized as follows. In Section \ref{isospin}, we
discuss some isospin-dependent bulk and single-particle properties
of asymmetric nuclear matter, such as the nuclear symmetry energy,
the nuclear symmetry potential, and the isospin-splitting of
nucleon effective mass as well as current experimental and/or
empirical constraints on these quantities. The theoretical
frameworks for the different versions of RMF models, i.e., the
nonlinear RMF models, the models with density-dependent
nucleon-meson coupling, and the nonlinear and density-dependent
point-coupling models, are briefly reviewed in Section \ref{RMF}.
Results on the isospin-dependent properties of asymmetric nuclear
matter, i.e., the nuclear symmetry energy, the nuclear symmetry
potential, and the isospin-splitting of nucleon effective mass and
the nucleon scalar densities in neutron-rich nuclear matter, from
different versions of RMF models are presented and discussed in
Section \ref{results}. A summary is then given in Section
\ref{summary}. For completeness, the isospin- and
momentum-dependent MDI interaction, which will be used as a
reference in some cases for comparison, is briefly described in
Appendix \ref{MDI}.

\section{Isospin-dependent properties of asymmetric nuclear matter}

\label{isospin}

\subsection{Nuclear symmetry energy}

The EOS of isospin asymmetric nuclear matter, given by its binding
energy per nucleon, can be generally written as
\begin{equation}
E(\rho ,\alpha )=E(\rho ,\alpha =0)+E_{\mathrm{sym}}(\rho )\alpha
^{2}+O(\alpha ^{4}),  \label{EsymPara}
\end{equation}%
where $\rho =\rho _{n}+\rho _{p}$ is the baryon density with $\rho
_{n}$ and $\rho _{p}$ denoting the neutron and proton densities,
respectively; $\alpha =(\rho _{n}-\rho _{p})/(\rho _{p}+\rho _{n})$
is the isospin asymmetry; $E(\rho ,\alpha =0)$ is the binding energy
per nucleon in symmetric nuclear matter, and
\begin{equation}
E_{\mathrm{sym}}(\rho )=\frac{1}{2}\frac{\partial ^{2}E(\rho ,\alpha
)}{\partial \alpha ^{2}}\vert_{\alpha =0}  \label{Esym}
\end{equation}
is the nuclear symmetry energy. The absence of odd-order terms in
$\alpha $ in Eq. (\ref{EsymPara}) is due to the exchange symmetry
between protons and neutrons in nuclear matter when one neglects the
Coulomb interaction and assumes the charge symmetry of nuclear
forces. The higher-order terms in $ \alpha$ are negligible, e.g.,
the magnitude of the $\alpha ^{4}$ term at $\rho _{0}$ is estimated
to be less than $1$ MeV \cite{Sie70,Sjo74,Lag81}. Neglecting the
contribution from higher-order terms in Eq. (\ref{EsymPara}) leads
to the well-known empirical parabolic law for the EOS of asymmetric
nuclear matter, which has been verified by all many-body theories to
date, at least for densities up to moderate values. As a good
approximation, the density-dependent symmetry energy
$E_{\mathrm{sym}}(\rho )$ can be extracted from
$E_{\mathrm{sym}}(\rho )\approx E(\rho ,\alpha =1)-E(\rho ,\alpha
=0)$, i.e., the energy change per nucleon when all protons in the
symmetric nuclear matter are converted to neutrons while keeping the
total nuclear density fixed. In this sense, the nuclear symmetry
energy gives an estimation of the binding energy difference between
the pure neutron matter without protons and the symmetric nuclear
matter with equal numbers of protons and neutrons. It should be
mentioned that the possible presence of the higher-order terms in $
\alpha$ at supra-normal densities can significantly modify the
proton fraction in $\beta$-equilibrium neutron-star matter and the
critical density for the direct Urca process which can lead to
faster cooling of neutron stars \cite{Zha01,Ste06}.

Around the nuclear matter saturation density $\rho _{0}$, the
nuclear symmetry energy $E_{\mathrm{sym}}(\rho )$\ can be expanded
to second-order in density as
\begin{equation}
E_{\mathrm{sym}}(\rho )=E_{\mathrm{sym}}(\rho
_{0})+\frac{L}{3}\left( \frac{\rho -\rho _{0}}{\rho _{0}}\right)
+\frac{K_{\mathrm{sym}}}{18}\left( \frac{\rho -\rho _{0}}{\rho
_{0}}\right) ^{2},  \label{EsymLK}
\end{equation}
where $L$ and $K_{\mathrm{sym}}$ are the slope and curvature
parameters of the nuclear symmetry energy at $\rho _{0}$, i.e.,
\begin{eqnarray}
L &=&3\rho _{0}\frac{\partial E_{\mathrm{sym}}(\rho )}{\partial \rho }
|_{\rho =\rho _{0}},  \label{L} \\
K_{\mathrm{sym}} &=&9\rho _{0}^{2}\frac{\partial
^{2}E_{\mathrm{sym}}(\rho )}{\partial ^{2}\rho }|_{\rho =\rho _{0}}.
\label{Ksym}
\end{eqnarray}
The $L$ and $K_{\mathrm{sym}}$ characterize the density dependence
of the nuclear symmetry energy around normal nuclear matter density,
and thus carry important information on the properties of nuclear
symmetry energy at both high and low densities. In particular, the
slope parameter $L$ has been found to correlate linearly with the
neutron-skin thickness of heavy nuclei and thus can in principle be
determined from measured thickness of the neutron skin of such
nuclei \cite{Bro00,Hor01a,Typ01,Fur02,Kar02,Die03,Che05b,Ste05b}.
Unfortunately, because of the large uncertainties in the
experimental measurements, this has not yet been possible so far.

At the nuclear matter saturation density and around $\alpha =0$,
the isobaric incompressibility of asymmetric nuclear matter can
also be expressed to second-order in $\alpha $ as
\cite{Pra85,Lop88}
\begin{equation}
K(\alpha )\approx K_{0}+K_{\mathrm{asy}}\alpha ^{2},  \label{Kasy}
\end{equation}
where $K_{0}$ is the incompressibility of symmetric nuclear matter
at the nuclear matter saturation density and the isospin-dependent
part \cite{Bar02}
\begin{equation}
K_{\mathrm{asy}}\approx K_{\mathrm{sym}}-6L  \label{Kasy2}
\end{equation}%
characterizes the density dependence of the nuclear symmetry energy.
Information on $K_{\mathrm{asy}}$ can in principle be extracted
experimentally by measuring the GMR in neutron-rich nuclei. Earlier
attempts based on this method have given, however, widely different
values. For example, a value of $K_{\mathrm{asy}}=-320\pm 180$ MeV
with a large uncertainty was obtained in Ref. \cite{Sha88} from a
systematic study of the GMR in the isotopic chains of Sn and Sm. In
this analysis, the value of $K_{0}$ was found to be $300\pm 25$ MeV,
which is somewhat larger than the commonly accepted value of $230\pm
10$ MeV. In a later study, an even less stringent constraint of
$-566\pm 1350<K_{\mathrm{asy}}<139\pm 1617$ MeV was extracted from
the GMR of finite nuclei, depending on the mass region of nuclei and
the number of parameters used in parameterizing the
incompressibility of finite nuclei \cite{Shl93}. Most recently, a
much stringent constraint of $K_{\mathrm{asy}}=-550\pm 100$ MeV has
been obtained in Ref. \cite{Gar07} from measurements of the isotopic
dependence of the GMR in even-A Sn isotopes.

Besides studies of nuclear structure, heavy-ion reactions,
especially those induced by radioactive beams, also provide a
useful means to investigate in terrestrial laboratories the
isospin-dependent properties of asymmetric nuclear matter,
particularly the density dependence of the nuclear symmetry
energy. Indeed, significant progress has recently been made both
experimentally and theoretically in extracting the information on
the behaviors of nuclear symmetry energy at sub-saturation density
from the isospin diffusion data in heavy-ion collisions from the
NSCL/MSU \cite{Tsa04,Che05a,LiBA05c}. Using the isospin and
momentum-dependent IBUU04 transport model with in-medium NN cross
sections, the isospin diffusion data were found to be consistent
with a density-dependent symmetry energy of $E_{\mathrm{sym}}(\rho
)\approx 31.6(\rho /\rho _{0})^{\gamma }$ with $\gamma =0.69-1.05$
at subnormal density \cite{Che05a,LiBA05c}, which has led to the
extraction of a value of $L=88\pm 25$ MeV for the slope parameter
of the nuclear symmetry energy at saturation density and a value
of $K_{\mathrm{asy}}=-500\pm 50$ MeV for the isospin-dependent
part of the isobaric incompressibility of isospin asymmetric
nuclear matter \cite{Che05a,LiBA05c,Che05b}. This has further
imposed stringent constraints on both the parameters in the
isospin-dependent nuclear effective interactions and the neutron
skin thickness of heavy nuclei. Among the $21$ sets of Skyrme
interactions commonly used in nuclear structure studies, only the
$4$ sets SIV, SV, G$_{\sigma }$, and R$_{\sigma }$ have been found
to give symmetry energies that are consistent with above extracted
one. Using these Skyrme interactions, the neutron-skin thickness
of heavy nuclei calculated within the Hartree-Fock approach is
consistent with available experimental data \cite{Che05b,Ste05b}
and also that from a relativistic mean-field model based on an
accurately calibrated parameter set that reproduces the GMR in
$^{90}$Zr and $^{208}$Pb as well as the isovector giant dipole
resonance of $^{208}$Pb \cite{Tod05}. The extracted symmetry
energy further agrees with the symmetry energy
$E_{\mathrm{sym}}(\rho )=31.6(\rho /\rho _{0})^{0.69}$ recently
obtained from the isoscaling analyses of isotope ratios in
intermediate energy heavy ion collisions \cite{She07}, which gives
$L\approx 65$ MeV and $K_{\mathrm{asy}}\approx -453$ MeV. The
extracted value of $K_{\mathrm{asy}}=-500\pm 50$ MeV from the
isospin diffusion data is also consistent with the value
$K_{\mathrm{asy}}=-550\pm 100$ MeV obtained from recently measured
isotopic dependence of the GMR in even-A Sn isotopes \cite{Gar07}.
We note that the GMR only allows us to extract the value of
$K_{\mathrm{asy}}$ but not that of $L$. These empirically
extracted values for $L$ and $K_{\rm sym}$ represent the best and
most stringent phenomenological constraints available so far on
the nuclear symmetry energy at sub-normal densities. Although the
behavior of the symmetry energy at high densities is presently
largely undetermined, much of this information is expected to be
obtained from future high energy radioactive beam facilities.

\subsection{Nuclear symmetry potential}

The nuclear symmetry potential refers to the isovector part of the
nucleon mean-field potential in isospin asymmetric nuclear matter.
Besides the nuclear density, the symmetry potential of a nucleon
in nuclear matter also depends on the momentum or energy of the
nucleon. The nuclear symmetry potential is different from the
nuclear symmetry energy as the latter involves the integration of
the isospin-dependent mean-field potential of a nucleon over its
momentum. The nuclear symmetry potential is thus a dynamical
quantity while the nuclear symmetry energy is a thermodynamic
quantity, and both are important for understanding many physics
questions in nuclear physics and astrophysics. Various microscopic
and phenomenological models have been used to study the symmetry
potential
\cite{Ulr97,Fuc04,Ma04,Sam05a,Fuc05,Fuc05b,Ron06,Bom91,Zuo05,Bar05,Das03,LiBA04a,LiBA04c,Che04,Riz04,Beh05,Riz05},
and the predicted results vary widely as in the case of the
nuclear symmetry energy. In particular, whereas most models
predict a decreasing symmetry potential with increasing nucleon
momentum albeit at different rates, a few nuclear effective
interactions used in some models give an opposite behavior.

The nuclear symmetry potential was originally defined in
non-relativistic models. In particular, the nuclear symmetry
potential can be evaluated from
\begin{equation}
U_{\mathrm{sym}}(\rho,\vec{p})=\frac{U_n(\rho,\vec{p})-U_p(\rho,\vec{p})}
{2\alpha}\label{Usym}
\end{equation}
where $U_n(\rho,\vec{p})$ and $U_p(\rho,\vec{p})$ represent,
respectively, the neutron and proton single-particle or mean-field
potentials In relativistic models, the nuclear symmetry potential
can be similarly defined by using the non-relativistic reduction
of the relativistic single-nucleon potentials. The nuclear
symmetry potential in relativistic models therefore depends on the
definition of the real part of the non-relativistic optical
potential or the nucleon mean-field potential deduced from the
relativistic effective interactions, which are characterized by
Lorentz covariant nucleon self-energies. In the relativistic
mean-field approximation, these self-energies appear in the
single-nucleon Dirac equation
\begin{equation}
\lbrack \gamma _{\mu }(i\partial ^{\mu }-\Sigma _{\tau }^{\mu
})-(M_{\tau }+\Sigma _{\tau }^{S})]\psi _{\tau }=0,~~~\tau=n,p
\label{Dirac}
\end{equation}
as the isospin-dependent nucleon vector self-energy $\Sigma _{\tau
}^{\mu }$ and scalar self-energy $\Sigma _{\tau }^{S}$. In the
Hartree approximation at the static limit, there are no currents
in a nucleus or nuclear matter, and the spatial vector components
vanish and only the time-like component of the vector self-energy
$\Sigma _{\tau }^{0}$ remains. Furthermore, the nucleon
self-energy is an energy-independent real and local quantity in
the standard RMF model.

There are different methods to derive the real part of the
non-relativistic optical potential based on the Dirac equation
with Lorentz covariant nucleon vector and scalar self-energies.
The most popular one is the so-called \textquotedblleft
Schr\"{o}dinger-equivalent potential" (SEP). From the nucleon
scalar self-energy $\Sigma _{\tau }^{S}$ and the time-like
component of the vector self-energy $\Sigma _{\tau }^{0}$, the
\textquotedblleft Schr\"{o}dinger-equivalent potential
\textquotedblright\ is given by \cite{Jam80}:
\begin{eqnarray}
U_{\mathrm{SEP},\tau } &=&\Sigma _{\tau }^{S}+\frac{1}{2M_{\tau }}[(\Sigma
_{\tau }^{S})^{2}-(\Sigma _{\tau }^{0})^{2}]+\frac{\Sigma _{\tau }^{0}}
{M_{\tau }}E_{\tau }  \notag \\
&=&\Sigma _{\tau }^{S}+\Sigma _{\tau }^{0}+\frac{1}{2M_{\tau
}}[(\Sigma _{\tau }^{S})^{2}-(\Sigma _{\tau }^{0})^{2}]
+\frac{\Sigma _{\tau }^{0}}{M_{\tau }}E_{\mathrm{kin}},\notag\\
\label{Usep}
\end{eqnarray}%
where $E_{\mathrm{kin}}$ is the kinetic energy of a nucleon, i.e.,
$E_{\mathrm{kin}}=E_{\tau }-M_{\tau }$ with $E_{\tau }$ being its
total energy. Eq.(\ref{Usep}) shows that $U_{\mathrm{SEP},\tau }$
increases linearly with the nucleon energy $E_{\tau }$ or kinetic
energy $E_{\mathrm{kin}}$ if the nucleon self-energies are
independent of energy. We note that by construction solving the
Schr\"{o}dinger equation with above SEP gives same bound-state
energy eigenvalues and elastic phase shifts as the solution of the
upper component of the Dirac spinor in the Dirac equation with
same nucleon scalar self-energy and time-like component of the
vector self-energy \cite{Jam80}. The above SEP thus best
represents the real part of the nucleon optical potential in
non-relativistic models \cite{Jam89,Fuc05}. The corresponding
nuclear symmetry potential is given by
\begin{equation}
U_{\mathrm{sym}}^{\mathrm{SEP}}=\frac{U_{\mathrm{SEP},n}-U_{\mathrm{SEP},p}}
{2\alpha},  \label{UsymSEP}
\end{equation}
with $\alpha $ being the isospin asymmetry.

Another popular alternative for deriving the non-relativistic
nucleon optical potential in relativistic models is to take it as
the difference between the total energy $E_{\tau }$ of a nucleon
with momentum $\vec{p}$ in the nuclear medium and its energy at the
same momentum in free space \cite{Fel91}, i.e.,
\begin{eqnarray}
U_{\mathrm{OPT},\tau } &=&E_{\tau }-\sqrt{\mathbf{p}^{2}+M_{\tau }^{2}}
\notag \\
&=&E_{\tau }-\sqrt{(E_{\tau }-\Sigma _{\tau }^{0})^{2}-\Sigma _{\tau
}^{S}(2M_{\tau }+\Sigma _{\tau }^{S})}.  \label{Uopt}
\end{eqnarray}
In obtaining the last step in above equation, the dispersion
relation
\begin{equation}
E_{\tau }=\Sigma _{\tau }^{0}+\sqrt{\mathbf{p}^{2}+(M_{\tau }+\Sigma _{\tau
}^{S})^{2}}  \label{dispersion}
\end{equation}%
has been used. This definition for the nucleon optical potential
has also been extensively used in microscopic DBHF calculations
\cite{LiGQ93} and transport models for heavy-ion collisions
\cite{Dan00}. For energy-independent nucleon self-energies,
$U_{\mathrm{OPT},\tau }$ approaches a constant value of $\Sigma
_{\tau }^{0}$ when $\left\vert \vec{p}\right\vert \rightarrow
\infty $, unlike the linear increase of $U_{\mathrm{SEP},\tau }$
with the nucleon energy. For $\left\vert \vec{p}\right\vert =0$,
we have $U_{\mathrm{OPT},\tau }=\Sigma _{\tau }^{S}+\Sigma _{\tau
}^{0}$ while $U_{\mathrm{SEP},\tau }=\Sigma _{\tau }^{S}+\Sigma
_{\tau }^{0}+(\Sigma _{\tau }^{S}+\Sigma _{\tau
}^{0})^{2}/(2M_{\tau })$. Therefore, $U_{\mathrm{OPT},\tau }$
displays a more reasonable high energy behavior than
$U_{\mathrm{SEP},\tau }$. We note that unlike
$U_{\mathrm{SEP},\tau }$, $U_{\mathrm{OPT},\tau }$ does not give
the same bound-state energy eigenvalues and elastic phase shifts
as the solution of the upper component of the Dirac equation. As
in the case of $U_{\mathrm{SEP},\tau }$, the symmetry potential in
this approach is defined by
\begin{equation}
U_{\mathrm{sym}}^{\mathrm{OPT}}=\frac{U_{\mathrm{OPT},n}-U_{\mathrm{OPT},p}}
{2\alpha}.  \label{UsymOPT}
\end{equation}

In Ref. \cite{Ham90}, another optical potential was introduced based
on the second-order Dirac (SOD) equation, and it corresponds to
multiplying Eq.(\ref{Usep}) by the factor $M_{\tau }/E_{\tau }$,
i.e.,
\begin{eqnarray}
U_{\mathrm{SOD},\tau } &=&[\Sigma _{\tau }^{S}+\frac{1}{2M_{\tau }}[(\Sigma
_{\tau }^{S})^{2}-(\Sigma _{\tau }^{0})^{2}]+\frac{\Sigma _{\tau }^{0}}
{M_{\tau }}E_{\tau }]\frac{M_{\tau }}{E_{\tau }}  \notag \\
&=&\Sigma _{\tau }^{0}+\frac{M_{\tau }}{E_{\tau }}\Sigma _{\tau
}^{S}+\frac{1}{2E_{\tau }}[(\Sigma _{\tau }^{S})^{2}-(\Sigma _{\tau
}^{0})^{2}]. \label{Usd}
\end{eqnarray}
For energy-independent nucleon self-energies, $U_{\text{SOD,}\tau
}$ has the same asymptotical value of $\Sigma _{\tau }^{0}$ as
$U_{\mathrm{OPT},\tau }$ when $\left\vert \vec{p}\right\vert
\rightarrow \infty $. For $\left\vert \vec{p}\right\vert =0$, we
have $U_{\mathrm{SOD},\tau }=\Sigma _{\tau }^{0}+\frac{M_{\tau
}}{\Sigma _{\tau }^{S}+\Sigma _{\tau }^{0}+M_{\tau }}\Sigma _{\tau
}^{S}+\frac{1}{2(\Sigma _{\tau }^{S}+\Sigma _{\tau }^{0}+M_{\tau
})} [(\Sigma _{\tau }^{S})^{2}-(\Sigma _{\tau }^{0})^{2}] $. The
symmetry potential based on the optical potential of Eq.
(\ref{Usd}) is given by
\begin{equation}
U_{\mathrm{sym}}^{\mathrm{SOD}}=\frac{U_{\mathrm{SOD},n}-U_{\mathrm{SOD},p}}
{2\alpha}.  \label{UsymSOD}
\end{equation}

Above discussions thus show that the optical potentials defined in
Eqs. (\ref{Uopt}) and (\ref{Usd}) have similar high energy
behaviors, but they may be very different from that defined in Eq.
(\ref{Usep}). If we assume that $\Sigma _{\tau }^{S}+\Sigma _{\tau
}^{0}\ll M_{\tau }$ and $\left\vert \Sigma _{\tau }^{S}\right\vert
\approx \left\vert \Sigma _{\tau }^{0}\right\vert $, which has
been shown to be generally valid in the RMF model even at higher
baryon densities, we have, however, $U_{\mathrm{SEP},\tau }\approx
U_{\mathrm{SOD},\tau }\approx U_{\mathrm{OPT},\tau }=\Sigma _{\tau
}^{S}+\Sigma _{\tau }^{0}$ at low momenta ($\left\vert
\vec{p}\right\vert \approx 0$), indicating that above three
definitions for the optical potential in the RMF model behave
similarly at low energies. However, it should be stressed that,
among the three optical potentials defined above, only
$U_{\mathrm{SEP},\tau }$ is obtained from a well-defined
theoretical procedure and is Schr\"{o}dinger-equivalent while
$U_{\mathrm{OPT},\tau }$ and $U_{\mathrm{SOD},\tau }$ are used
here for heuristic reasons as they are of practical interest in
microscopic DBHF calculations, transport models for heavy-ion
collisions, and Dirac phenomenology.

Empirically, a systematic analysis of a large number of
nucleon-nucleus scattering experiments and (p,n) charge-exchange
reactions at beam energies up to about $100$ MeV has shown that
the data can be very well described by the parametrization
$U_{\mathrm{sym}}=a-bE_{\mathrm{kin}}$ with $a\approx 22-34$ MeV
and $b\approx 0.1-0.2$ \cite{Sat69,Hof72,Hod94,Kon03}. Although
the uncertainties in both parameters $a$ and $b$ are large, the
nuclear symmetry potential at nuclear matter saturation density,
i.e., the Lane potential $U_{\mathrm{Lane}}$ \cite{Lan62}, clearly
decreases approximately linearly with increasing beam energy
$E_{\mathrm{kin}}$. This provides a stringent constraint on the
low energy behavior of the nuclear symmetry potential at
saturation density. As we will see in the following, although the
predicted energy dependence of nuclear symmetry potential at low
energy from the RMF models does not agree with the empirical Lane
potential, it is consistent with results from microscopic DBHF
\cite{Fuc04}, the extended BHF with 3-body forces \cite{Zuo05},
and chiral perturbation theory calculations \cite{Fri05}, which
give a Lane potential that also stays as a constant or increases
slightly with momentum for nucleons with momenta less than about
$250-300$ MeV/c or with kinetic energies $E_{\mathrm{kin}}<0$ but
decreases with momentum when the momentum is larger than about
$250-300$ MeV/c.

Recently, the high energy behavior of the nuclear symmetry
potential has been studied in the relativistic impulse ($t$-$\rho
$) approximation based on the empirical NN scattering amplitude
\cite{Che05c}. The results indicate that the nuclear symmetry
potential derived from the Schr\"{o}dinger-equivalent potential at
a fixed density becomes almost constant when the nucleon kinetic
energy is greater than about $500$ MeV, independent of the
parameters used in the analysis. It is further shown that for such
high energy nucleons the nuclear symmetry potential is slightly
negative at baryon densities below about $\rho =0.22$ fm$^{-3}$
and then increases almost linearly to positive values at high
densities. These results provide important constraints on the high
energy behavior of the nuclear symmetry potential in asymmetric
nuclear matter. Furthermore, with the Love-Franey NN scattering
amplitude developed by Murdock and Horowitz \cite{Hor85,Mur87},
the intermediate-energy ($100\leq $ $E_{kin}\leq 400$ MeV)
behavior of the nuclear symmetry potential constructed from the
Schr\"{o}dinger-equivalent potential in isospin asymmetric nuclear
matter has also been investigated recently \cite{LiZH06b}. It
shows that the nuclear symmetry potential at fixed baryon density
decreases with increasing nucleon energy. In particular, the
nuclear symmetry potential at saturation density changes from
positive to negative values at nucleon kinetic energy of about
$200$ MeV. Such an energy and density dependence of the nuclear
symmetry potential is consistent with those from the isospin- and
momentum-dependent MDI interaction with $x=0$ (see Appendix
\ref{MDI} for details on this interaction). These results thus
provide an important consistency check for the energy/momentum
dependence of the nuclear symmetry potential in asymmetric nuclear
matter, particularly the MDI interaction with $x=0$. On the other
other, the low energy behavior of the nuclear symmetry potential
at densities away from normal nuclear density is presently not
known empirically. Experimental determination of both the density
and momentum dependence of the nuclear symmetry potential is thus
of great interest, and heavy-ion reactions with radioactive beams
provides a unique tool to extract this information in terrestrial
laboratories.

\subsection{Nucleon effective mass}

Many different definitions for the nucleon effective mass can be
found in the literature \cite{Jam89,Fuc05}. In the present work,
we mainly focus on the following three effective masses: the Dirac
mass $M_{\mathrm{Dirac}}^{\ast }$ (also denoted as $M^{\ast }$ in
the present work), the Landau mass $M_{\mathrm{Landau}}^{\ast }$,
and the Lorentz mass $M_{\mathrm{Lorentz}}^{\ast } $. The Dirac
mass $M_{\mathrm{Dirac}}^{\ast }$ is defined through the nucleon
scalar self-energy in the Dirac equation, i.e.,
\begin{equation}
M_{\mathrm{Dirac},\tau }^{\ast }=M_{\tau }+\Sigma _{\tau }^{S}.
\end{equation}
It is directly related to the spin-orbit potential in finite nuclei
and is thus a genuine relativistic quantity without non-relativistic
correspondence. We note that the difference between the nucleon
vector and scalar self-energies determines the spin-orbit potential,
whereas their sum defines the effective single-nucleon potential and
is constrained by the nuclear matter binding energy at saturation
density. From the energy spacings between spin-orbit partner states
in finite nuclei, the constraint $0.55~M$ $\leq
M_{\mathrm{Dirac}}^{\ast }\leq 0.6~M$ has been obtained on the value
of the Dirac mass \cite{Typ05,Mar07}.

The Landau mass $M_{\mathrm{Landau}}^{\ast }$ is defined as
$M_{\mathrm{Landau},\tau }^{\ast } =p\frac{dp}{dE_{\tau }}$ in terms
of the single-particle density of state $dE_{\tau }/dp$ at energy
$E_{\tau }$ and thus characterizes the momentum dependence of the
single-particle potential. In the relativistic model, it is given by
\cite{Typ05}
\begin{eqnarray}
M_{\mathrm{Landau},\tau }^{\ast } =(E_{\tau }-\Sigma _{\tau
}^{0})(1- \frac{d\Sigma_{\tau}^{0}}{dE_{\tau }})-(M_{\tau }+\Sigma
_{\tau }^{S})\frac{d\Sigma _{\tau }^{S}}{dE_{\tau }}.
\label{MLandau}
\end{eqnarray}
Since $dp/dE_{\tau }$ is in principle measurable, the Landau mass
from the relativistic model should have a comparable value as that
in the non-relativistic model. Empirically, based on
non-relativistic effective interactions such as the Skyrme-type
interactions, calculations of the ground-state properties and the
excitation energies of quadrupole giant resonances have shown that
a realistic choice for the nucleon Landau mass is
$M_{\mathrm{Landau}}^{\ast }/M$ = $0.8\pm 0.1$
\cite{Cha97,Cha98,Rei99,Mar07}. The smaller Landau mass than that
of nucleon free mass would lead to a smaller level density at the
Fermi energy and much spreaded single-particle levels in finite
nuclei \cite{Typ05}.

The Lorentz mass $M_{\mathrm{Lorentz}}^{\ast }$ characterizes the
energy dependence of the Schr\"{o}dinger-equivalent Potential
$U_{\mathrm{SEP},\tau }$ in the relativistic model and is defined as
\cite{Jam89}
\begin{eqnarray}
M_{\mathrm{Lorentz},\tau }^{\ast } &=&M_{\tau }(1-\frac{dU_{\mathrm{SEP},\tau }}
{dE_{\tau }})  \notag \\
&=&(E_{\tau }-\Sigma _{\tau }^{0})(1-\frac{d\Sigma _{\tau }^{0}}{dE_{\tau }})
\notag \\
&&-(M_{\tau }+\Sigma _{\tau }^{S})\frac{d\Sigma _{\tau }^{S}}{dE_{\tau }}
+M_{\tau }-E_{\tau }  \notag \\
&=&M_{\mathrm{Landau},\tau }^{\ast }+M_{\tau }-E_{\tau }.  \label{MLorentz}
\end{eqnarray}
It has been argued in Ref. \cite{Jam89} that it is the Lorentz mass
$M_{\mathrm{Lorentz}}^{\ast }$ that should be compared with the
usual non-relativistic nucleon effective mass extracted from
analyses carried out in the framework of non-relativistic optical
and shell models. It can be easily seen that in the non-relativistic
approximation ($E_{\tau }\approx M_{\tau }$), the Lorentz mass
$M_{\mathrm{Lorentz}}^{\ast }$ reduces to the Landau mass
$M_{\mathrm{Landau}}^{\ast } $.

In relativistic models, the nucleon effective mass has sometimes
also been introduced via the energy dependence of the optical
potential in Eq. (\ref{Uopt}) and the second-order Dirac optical
potential in Eq. (\ref{Usd}), i.e.,
\begin{eqnarray}
M_{\mathrm{OPT},\tau }^{\ast } &=&M_{\tau }(1-\frac{dU_{\mathrm{OPT},\tau }}
{dE_{\tau }})  \notag \\
&=&M_{\tau }\frac{(E_{\tau }-\Sigma _{\tau }^{0})(1-\frac{d\Sigma
_{\tau}^{0}} {dE_{\tau }})+(M_{\tau }-\Sigma _{\tau
}^{S})\frac{d\Sigma _{\tau }^{S}}{dE_{\tau }}}{\sqrt{(E_{\tau
}-\Sigma _{\tau }^{0})^{2}-\Sigma _{\tau
}^{S}(2M_{\tau }+\Sigma _{\tau }^{S})}}  \notag \\
&=&M_{\tau }\frac{M_{\mathrm{Landau},\tau }^{\ast }}{\sqrt{(E_{\tau }-\Sigma
_{\tau }^{0})^{2}-\Sigma _{\tau }^{S}(2M_{\tau }+\Sigma _{\tau }^{S})}}
\label{Mopt}
\end{eqnarray}
and
\begin{eqnarray}
&&M_{\mathrm{SOD},\tau }^{\ast }  \notag \\
&=&M_{\tau }(1-\frac{dU_{\mathrm{SOD},\tau }}{dE_{\tau }})  \notag \\
&=&M_{\tau }[\frac{M_{\mathrm{Landau},\tau }^{\ast }}{E_{\tau }}  \notag \\
&&+\frac{(M_{\tau }+\Sigma _{\tau }^{S})^{2}-(E_{\tau }-\Sigma _{\tau
}^{0})^{2}+E_{\tau }^{2}-M_{\tau }^{2}}{2E_{\tau }^{2}}],  \label{Msod}
\end{eqnarray}
respectively.

The isospin-splitting of nucleon effective mass in asymmetric
nuclear matter, i.e., the difference between the neutron and
proton effective masses is currently not known empirically
\cite{Lun03}. Previous theoretical investigations have indicated
that most RMF calculations with the isovector $\delta $ meson
predict $M_{\mathrm{Dirac},n}^{\ast }<M_{\mathrm{Dirac},p}^{\ast
}$ while in the microscopic DBHF approach,
$M_{\mathrm{Dirac},n}^{\ast }$ can be larger or smaller than
$M_{\mathrm{Dirac},p}^{\ast }$ depending on the approximation
schemes and methods used for determining the Lorentz and isovector
structure of the nucleon self-energy \cite{Fuc05}. For the nucleon
Lorentz mass, the microscopic DBHF or BHF approach and most
non-relativistic Skyrme-Hartree-Fock calculations predict $
M_{\mathrm{Lorentz},n}^{\ast }>M_{\mathrm{Lorentz},p}^{\ast }$,
while most RMF and a few Skyrme-Hartree-Fock calculations give
opposite predictions.

\section{Relativistic mean-field models}

\label{RMF}

For completeness, we briefly introduce in the following the main
ingredients in the nonlinear RMF model, the density-dependent RMF
model, the nonlinear point-coupling RMF model, and the
density-dependent point-coupling RMF model. We neglect the
electromagnetic field in the following since in the present work
we are interested in the properties of the infinite nuclear
matter. Furthermore, besides the mean-field approximation in which
operators of meson fields are replaced by their expectation values
(the fields are thus treated as classical c-numbers), we also use
the non-sea approximation which neglects the effect due to
negative energy states in the Dirac sea.

\subsection{The nonlinear RMF model}

\subsubsection{Lagrangian density}

The Lagrangian density in the nonlinear RMF model generally includes
the nucleon field $\psi $, the isoscalar-scalar meson field $\sigma
$, the isoscalar-vector meson field $\omega $, the isovector-vector
meson field $\vec{\rho}$, and the isovector-scalar meson field
$\delta $, i.e.,
\begin{eqnarray}
&&\mathcal{L}_{\mathrm{NL}}=\bar{\psi}\left[ \gamma _{\mu }(i\partial ^{\mu
}-g_{\omega }\omega ^{\mu })-(M-g_{\sigma }\sigma )\right] \psi  \notag \\
&&+\frac{1}{2}(\partial _{\mu }\sigma \partial ^{\mu }\sigma -m_{\sigma
}^{2}\sigma ^{2})-\frac{1}{4}\omega _{\mu \nu }\omega ^{\mu \nu }+\frac{1}{2}
m_{\omega }^{2}\omega _{\mu }\omega ^{\mu }  \notag \\
&&-\frac{1}{3}b_{\sigma }M{(g_{\sigma }\sigma )}^{3}-\frac{1}{4}c_{\sigma }{%
\ (g_{\sigma }\sigma )}^{4}+\frac{1}{4}c_{\omega }{(g_{\omega }^{2}\omega
_{\mu }\omega ^{\mu })}^{2}  \notag \\
&&+\frac{1}{2}(\partial _{\mu }\vec{\delta}\cdot \partial ^{\mu }\vec{\delta}
-m_{\delta }^{2}\vec{\delta}^{2})+\frac{1}{2}m_{\rho }^{2}\vec{\rho}_{\mu
}\cdot \vec{\rho}^{\mu }-\frac{1}{4}\vec{\rho}_{\mu \nu }\cdot \vec{\rho}
^{\mu \nu }  \notag \\
&&+\frac{1}{2}(g_{\rho }^{2}\vec{\rho}_{\mu }\cdot \vec{\rho}^{\mu
})(\Lambda _{S}g_{\sigma }^{2}\sigma ^{2}+\Lambda _{V}g_{\omega }^{2}{\omega
_{\mu }\omega ^{\mu }})  \notag \\
&&-g_{\rho }\vec{\rho}_{\mu }\cdot \bar{\psi}\gamma ^{\mu }\vec{\tau}\psi
+g_{\delta }\vec{\delta}\cdot \bar{\psi}\vec{\tau}\psi \;,  \label{lagNL}
\end{eqnarray}
where the antisymmetric field tensors $\omega _{\mu \nu }$ and
$\vec{\rho} _{\mu \nu }$ are given by $\omega _{\mu \nu }\equiv
\partial _{\nu }\omega _{\mu }-\partial _{\mu }\omega _{\nu }$ and
$\text{ }\vec{\rho}_{\mu \nu }\equiv \partial _{\nu
}\vec{\rho}_{\mu }-\partial _{\mu }\vec{\rho}_{\nu }$ ,
respectively, and other symbols have their usual meanings. Also,
vectors in isospin space are denoted by arrows. This model also
contains cross interactions between the isovector meson $\rho $
and isoscalar $\sigma $ and $\omega $ mesons through the
cross-coupling constants $\Lambda _{S}$ and $\Lambda _{V}$
\cite{Mul96,Hor01a}. In addition, we include the isovector-scalar
channel ($\delta $ meson) which is important for the saturation of
asymmetric nuclear matter and has also been shown to be an
important degree of freedom in describing the properties of
asymmetric nuclear matter \cite{Kub97,Liu02}. The above Lagrangian
density is quite general and allows us to use most of presently
popular parameter sets in the nonlinear RMF model.

\subsubsection{Equation of motion and nucleon self-energies}

From the standard Euler-Lagrange formalism, we can deduce from the
Lagrangian density equations of motion for the nucleon and meson
fields. The resulting Dirac equation for the nucleon field is
\begin{equation}
\left[ \gamma _{\mu }(i\partial ^{\mu }-\Sigma _{\tau }^{\mu
})-(M+\Sigma _{\tau }^{S})\right] \psi =0\;,
\end{equation}
with the following nucleon scalar and vector self-energies:
\begin{eqnarray}
\Sigma _{\tau }^{S}&=&-g_{\sigma }\sigma -g_{\delta
}\vec{\delta}\cdot \vec{\tau}, \\
\Sigma _{\tau }^{\mu }&=&g_{\omega }\omega ^{\mu }+g_{\rho
}\vec{\rho}^{\mu }\cdot \vec{\tau}.
\end{eqnarray}
%\begin{equation}
%\Sigma _{\tau }^{\mu }=g_{\omega }\omega ^{\mu }+g_{\rho
%}\vec{\rho}^{\mu }\cdot \vec{\tau}.
%\end{equation}

For the isoscalar meson fields $\sigma $ and $\omega $, they are
described by the Klein-Gordon and Proca equations, respectively,
i.e.,
\begin{eqnarray}
(\partial _{\mu }\partial ^{\mu }+m_{\sigma }^{2})\sigma &=&g_{\sigma }[\bar{
\psi}\psi -b_{\sigma }M{(g_{\sigma }\sigma )}^{2}-c_{\sigma }{%
(g_{\sigma}\sigma )}^{3}  \notag \\
&&+\Lambda _{S}{(g_{\sigma }\sigma )}g_{\rho }^{2}\vec{\rho}_{\mu }\cdot
\vec{\rho}^{\mu }]\;, \\
\partial _{\mu }\omega ^{\mu \nu }+m_{\omega }^{2}\omega ^{\nu }
&=&g_{\omega }[\bar{\psi}\gamma ^{\nu }\psi -c_{\omega }g_{\omega
}^{3}(\omega _{\mu }\omega ^{\mu }\omega ^{\nu })  \notag \\
&&-\Lambda _{V}g_{\rho }^{2}\vec{\rho}_{\mu }\cdot \vec{\rho}^{\mu
}g_{\omega }\omega ^{\nu }]\;.
\end{eqnarray}

Analogous equations for the isovector $\delta $ and $\rho $ meson fields are
\begin{eqnarray}
(\partial _{\mu }\partial ^{\mu }+m_{\delta }^{2})\vec{\delta} &=&g_{\delta
} \bar{\psi}\vec{\tau}\psi , \\
\partial _{\mu }\vec{\rho}^{\mu \nu }+m_{\rho }^{2}\vec{\rho}^{\nu }
&=&g_{\rho }[\bar{\psi}\gamma ^{\nu }\vec{\tau}\psi -\Lambda _{S}(g_{\rho }%
\vec{\rho}^{\nu }){(g_{\sigma }\sigma )}^{2}  \notag \\
&&-\Lambda _{V}(g_{\rho }\vec{\rho}^{\nu })g_{\omega }^{2}\omega _{\mu
}\omega ^{\mu }].
\end{eqnarray}

For a static, homogenous infinite nuclear matter, all derivative
terms drop out and the expectation values of space-like components
of vector fields vanish (only zero components $\vec{\rho}_{0}$ and
$\omega _{0}$ survive) due to translational invariance and
rotational symmetry of the nuclear matter. In addition, only the
third component of isovector fields ($\delta ^{(3)}$ and $\rho
^{(3)}$) needs to be taken into consideration due to the
rotational invariance around the third axis in the isospin space.
In the mean-field approximation, meson fields are replaced by
their expectation values, i.e., $\sigma \rightarrow \bar{\sigma}$,
$\omega _{\mu }\rightarrow \bar{\omega}_{0}$,
$\vec{\delta}\rightarrow \bar{\delta}^{(3)}$, and $\vec{
\rho}_{\mu }\rightarrow \bar{\rho}_{0}^{(3)}$, and the meson field
equations are reduced to
\begin{eqnarray}
m_{\sigma }^{2}\bar{\sigma} &=&g_{\sigma }[\rho _{S}-b_{\sigma }M{(g_{\sigma
}\bar{\sigma})}^{2}-c_{\sigma }{(g_{\sigma }\bar{\sigma})}^{3}  \notag \\
&&+\Lambda _{S}{(g_{\sigma }\bar{\sigma})(}g_{\rho }\bar{\rho}
_{0}^{(3)})^{2}], \\
m_{\omega }^{2}\bar{\omega}_{0} &=&g_{\omega }[\rho _{B}-c_{\omega }{%
(g_{\omega }\bar{\omega}_{0})}^{3}  \notag \\
&&-\Lambda {(g_{\omega }\bar{\omega}_{0})(}g_{\rho }\bar{\rho}
_{0}^{(3)})^{2}],  \label{OmgNL} \\
m_{\delta}^{2}{\bar{\delta}}^{(3)} &=&g_{\delta }(\rho _{S,p}-\rho _{S,n}).
\label{DelNL} \\
m_{\rho }^{2}\bar{\rho}_{0}^{(3)} &=&g_{\rho }[\rho _{B,p}-\rho
_{B,n}-\Lambda _{S}{(}g_{\rho }\bar{\rho}_{0}^{(3)}){(g_{\sigma }\sigma )}%
^{2}  \notag \\
&&-\Lambda _{V}{(}g_{\rho }\bar{\rho}_{0}^{(3)}){(g_{\omega }\bar{\omega}%
_{0})}^{{2}}].  \label{RhoNL}
\end{eqnarray}
In the above, the nucleon scalar density $\rho _{S}$ is defined as
\begin{equation}
\rho _{S}=\left\langle \bar{\psi}\psi \right\rangle =\rho _{S,p}+\rho
_{S,n}\;,  \label{RhoS}
\end{equation}
with the proton ($p$) and neutron ($n$) scalar densities given by
\begin{eqnarray}
\rho _{S,i} &=&\frac{2}{{(2\pi )}^{3}}\int_{0}^{k_{F}^{i}}d^{3}\!k\,\frac{%
M_{i}^{\ast }}{\sqrt{\vec{k}^{2}+(M_{i}^{\ast })^{2}}}  \notag \\
&=&\frac{M_{i}^{\ast }}{2\pi ^{2}}\left[ k_{F}^{i}\tilde{E}
_{F}^{i}-(M_{i}^{\ast })^{2}\ln \frac{k_{F}^{i}+\tilde{E}_{F}^{i}}{%
M_{i}^{\ast }}\right] ,i=p,n \notag\\
\label{RhoSnp}
\end{eqnarray}
where
\begin{equation}
\tilde{E}_{F}^{i}=\sqrt{(k_{F}^{i})^{2}+(M_{i}^{\ast })^{2}},  \label{Ef}
\end{equation}
with $M_{p}^{\ast }$ and $M_{n}^{\ast }$ denoting the proton and
neutron Dirac masses, respectively, i.e.,
\begin{equation}
M_{p}^{\ast }=M-g_{\sigma }\bar{\sigma}-g_{\delta }{\bar{\delta}}^{(3)},
\text{ }M_{n}^{\ast }=M-g_{\sigma }\bar{\sigma}+g_{\delta }{\bar{\delta}}%
^{(3)}.  \label{MDiracNL}
\end{equation}
The nucleon scalar and vector self-energies are then given by
\begin{eqnarray}
\Sigma _{\tau }^{S} &=&-g_{\sigma }\bar{\sigma}-g_{\delta }{\bar{\delta}}
^{(3)}\tau _{3}, \\
\Sigma _{\tau }^{0} &=&g_{\omega }\bar{\omega}_{0}+g_{\rho }\bar{\rho}
_{0}^{(3)}\tau _{3},  \label{Sig0NL}
\end{eqnarray}
with $\tau _{3}=1$ and $-1$ for protons and neutrons, respectively.

\subsubsection{Nuclear matter equation of state}

The set of coupled equations for the nucleon and meson fields can
be solved self-consistently using the iteration method, and the
properties of the nuclear matter can then be obtained from these
fields. From the resulting energy-momentum tensor, we can
calculate the energy density $\epsilon $ and pressure $P$ of
asymmetric nuclear matter, and the results are given by
\begin{eqnarray}
\epsilon &=&\epsilon _{kin}^{n}+\epsilon _{kin}^{p}  \notag \\
&&+\frac{1}{2}\left[ m_{\sigma }^{2}\bar{\sigma}^{2}+m_{\omega }^{2}\bar{%
\omega}_{0}^{2}+m_{\delta }^{2} {\bar{\delta}}^{(3)2}+m_{\rho }^{2}\bar{\rho}
_{0}^{(3)2}\right]  \notag \\
&&+\frac{1}{3}b_{\sigma }M{(g_{\sigma }\bar{\sigma})}^{3}+\frac{1}{4}%
c_{\sigma }{(g_{\sigma }\bar{\sigma})}^{4}+\frac{3}{4}c_{\omega }{(g_{\omega
}\bar{\omega}_{0})}^{4}  \notag \\
&&+\frac{1}{2}(g_{\rho }\bar{\rho}_{0}^{(3)})^{2}[\Lambda _{S}{(g_{\sigma }%
\bar{\sigma})}^{2}+3\Lambda _{V}{(g_{\omega }\bar{\omega}_{0})}^{2}]
\end{eqnarray}
and
\begin{eqnarray}
P &=&P_{kin}^{n}+P_{kin}^{p}  \notag \\
&&-\frac{1}{2}\left[ m_{\sigma }^{2}\bar{\sigma}^{2}-m_{\omega }^{2} \bar{%
\omega}_{0}^{2}+m_{\delta }^{2}{\bar{\delta}}^{(3)2}-m_{\rho }^{2}\bar{\rho}%
_{0}^{(3)2}\right]  \notag \\
&&-\frac{1}{3}b_{\sigma }M{(g_{\sigma }\bar{\sigma})}^{3}-\frac{1}{4}
c_{\sigma }{(g_{\sigma }\bar{\sigma})}^{4}+\frac{1}{4}c_{\omega }{(g_{\omega
}\bar{\omega}_{0})}^{4}  \notag \\
&&+\frac{1}{2}(g_{\rho }\bar{\rho}_{0}^{(3)})^{2}[\Lambda _{S}{(g_{\sigma }
\bar{\sigma})}^{2}+\Lambda _{V}{(g_{\omega }\bar{\omega}_{0})}^{2}].
\end{eqnarray}
In the above, $\epsilon _{kin}^{i}$ and $P_{kin}^{i}$ are,
respectively, the kinetic contributions to the energy densities and
pressure of protons and neutrons in nuclear matter, and they are
given by
\begin{eqnarray}
\epsilon _{kin}^{i} &=&\frac{2}{(2\pi )^{3}}\int_{0}^{k_{F}^{i}}d^{3}k \sqrt{%
\vec{k}^{2}+(M_{i}^{\ast })^{2}}  \notag \\
&=&\frac{1}{4}[3\tilde{E}_{F}^{i}\rho _{B,i}+M_{i}^{\ast }\rho _{S,i}],\quad
i=p,n,
\end{eqnarray}
and
\begin{eqnarray}
P_{kin}^{i} &=&\frac{2}{3(2\pi )^{3}}\int_{0}^{k_{F}^{i}}d^{3}k\frac{\vec{k}
^{2}}{\sqrt{\vec{k}^{2}+(M_{i}^{\ast })^{2}}}  \notag \\
&=&\frac{1}{4}[\tilde{E}_{F}^{i}\rho _{B,i}-M_{i}^{\ast }\rho _{S,i}],\quad
i=p,n.
\end{eqnarray}

The binding energy per nucleon can be obtained from the energy density via
\begin{equation*}
E=\frac{\epsilon }{\rho _{B}}-M,
\end{equation*}
while the symmetry energy is given by
\begin{eqnarray}
E_{\mathrm{sym}}(\rho _{B}) &=&\frac{k_{F}^{2}}{6\tilde{E}_{F}}+\frac{1}{2}
\left( \frac{g_{\rho }}{{m_{\rho }^{\ast }}}\right) ^{2}\rho _{B}-\frac{1}{2}
\left( \frac{g_{\delta }}{m_{\delta }}\right) ^{2}  \notag \\
&&\times \frac{M^{\ast 2}\rho _{B}}{\tilde{E}_{F}^{2}[1+\left( \frac{
g_{\delta }}{m_{\delta }}\right) ^{2}A(k_{F},M^{\ast })]},  \label{EsymNL}
\end{eqnarray}
with the effective $\rho $-meson mass given by \cite{Hor01a}
\begin{equation}
{m_{\rho }^{\ast }}^{2}=m_{\rho }^{2}+g_{\rho }^{2}[\Lambda _{S}{(g_{\sigma
} \bar{\sigma})}^{2}+\Lambda _{V}{(g_{\omega }\bar{\omega}_{0})}^{2}]
\end{equation}
and
\begin{eqnarray}
A(k_{F},M^{\ast }) &=&\frac{4}{(2\pi )^{3}}\int_{0}^{k_{F}}d^{3}k\frac{\vec{%
k }^{2}}{\left( \vec{k}^{2}+(M^{\ast })^{2}\right) ^{3/2}}  \notag \\
&=&3\left( \frac{\rho _{S}}{M^{\ast }}-\frac{\rho _{B}}{\tilde{E}_{F}}
\right),
\end{eqnarray}
where $\tilde{E}_{F}=\sqrt{k_{F}^{2}+M^{\ast }{}^{2}}$ and $M^{\ast } $ is
the nucleon Dirac mass in symmetric nuclear matter.

\subsection{The density-dependent RMF model}

\subsubsection{Lagrangian density}

In the density-dependent RMF model, instead of introducing terms
involving self-interactions of the scalar meson field and
cross-interactions of meson fields as in the nonlinear RMF model,
the coupling constants are density dependent. The Lagrangian
density in this model is generally written as
\begin{eqnarray}
\mathcal{L}_{\mathrm{DD}} &=&\bar{\psi}[\gamma _{\mu }(i\partial ^{\mu
}-\Gamma _{\omega }\omega ^{\mu }-\Gamma _{\rho }\vec{\rho}^{\mu }\cdot \vec{%
\tau})  \notag \\
&&-(M-\Gamma _{\sigma }\sigma -\Gamma _{\delta }\vec{\delta}\cdot \vec{\tau}%
)]\psi  \notag \\
&&+\frac{1}{2}(\partial _{\mu }\sigma \partial ^{\mu }\sigma
-m_{s}^{2}\sigma ^{2})+\frac{1}{2}(\partial _{\mu }\vec{\delta}\cdot
\partial ^{\mu }\vec{\delta}-m_{\delta }^{2}\vec{\delta}^{2})  \notag \\
&&-\frac{1}{4}\omega _{\mu \nu }\omega ^{\mu \nu }+\frac{1}{2}m_{\omega
}^{2}\omega _{\mu }\omega ^{\mu }  \notag \\
&&-\frac{1}{4}\vec{\rho}_{\mu \nu }\cdot \vec{\rho}^{\mu \nu }+\frac{1}{2}
m_{\rho }^{2}\vec{\rho}_{\mu }\cdot \vec{\rho}^{\mu }  \label{lagDD}
\end{eqnarray}
The symbols used in above equation have their usual meanings as in
Eq.(\ref{lagNL}) but the coupling constants $\Gamma _{\sigma }$,
$\Gamma _{\omega }$, $\Gamma _{\delta }$ and $\Gamma _{\rho }$ now
depend on the (baryon) density, which are usually parametrized as
\begin{equation}
\Gamma _{i}(\rho )=\Gamma _{i}(\rho _{sat})h_{i}(x),\quad x=\rho /\rho
_{sat},
\end{equation}
with
\begin{equation}
h_{i}(x)=a_{i}\frac{1+b_{i}(x+d_{i})^{2}}{1+c_{i}(x+e_{i})^{2}},\quad
i=\sigma ,\omega ,\delta ,\rho ,
\end{equation}
and $\rho _{sat}$ being the saturation density of symmetric
nuclear matter. In some parameter sets,
\begin{equation} h_{\rho }(x)=\exp
[-a_{\rho }(x-1)]
\end{equation}
is used for the $\rho $ meson.

\subsubsection{Equation of motion and nucleon self-energies}

Since the coupling constants in the density-dependent RMF model
depend on the baryon fields $\bar{\psi}$ and $\psi $ through the
density, additional terms besides the usual ones in the nonlinear
RMF model appear in the field equations of motion when the partial
derivatives of $\mathcal{L}_{\text{DD}}$ are carried out with
respect to the fields $\bar{\psi}$ and $\psi $ in the
Euler-Lagrange equations. The resulting Dirac equation for the
nucleon field now reads:
\begin{equation}
\left[ \gamma _{\mu }(i\partial ^{\mu }-\Sigma _{\tau }^{\mu
})-(M+\Sigma _{\tau }^{S})\right] \psi =0,
\end{equation}
with the following nucleon scalar and vector self-energies:
\begin{eqnarray}
\Sigma _{\tau }^{S}&=&-\Gamma _{\sigma }\sigma -\Gamma _{\delta
}\vec{\delta} \cdot \vec{\tau},\\
\Sigma _{\tau }^{\mu }&=&\Gamma _{\omega }\omega ^{\mu }+\Gamma
_{\rho }\vec{ \rho}^{\mu }\cdot \vec{\tau}+\Sigma ^{\mu (R)}.
\end{eqnarray}
The new term $\Sigma ^{\mu (R)}$ in the vector self-energy, which
is called the \textit{rearrangement} self-energy
\cite{Len95,Fuc95}, is given by
\begin{eqnarray}
\Sigma ^{\mu (R)} &=&\frac{j^{\mu }}{\rho }(\frac{\partial \Gamma _{\omega }
}{\partial \rho }\bar{\psi}\gamma _{\nu }\psi \omega ^{\nu }+\frac{\partial
\Gamma _{\rho }}{\partial \rho }\bar{\psi}\vec{\tau}\gamma ^{\nu }\psi \cdot
\vec{\rho}_{\nu }  \notag \\
&&-\frac{\partial \Gamma _{\sigma }}{\partial \rho }\bar{\psi}\psi \sigma -
\frac{\partial \Gamma _{\delta }}{\partial \rho }\bar{\psi}\vec{\tau}\psi
\vec{\delta})~,
\end{eqnarray}
with $j^{\mu }=\bar{\psi}\gamma ^{\mu }\psi $. The rearrangement
self-energy plays an essential role in the applications of the
theory since it guarantees both the thermodynamical consistency
and the energy-momentum conservation \cite{Len95,Fuc95}.

For the meson fields, the equations of motion are
\begin{eqnarray}
(\partial _{\mu }\partial ^{\mu }+m_{\sigma }^{2})\sigma &=&\Gamma _{\sigma }%
\bar{\psi}\psi , \\
\partial _{\nu }\omega ^{\mu \nu }+m_{\omega }^{2}\omega ^{\mu } &=&\Gamma
_{\omega }\bar{\psi}\gamma ^{\mu }\psi , \\
(\partial _{\mu }\partial ^{\mu }+m_{\delta }^{2})\vec{\delta} &=&\Gamma
_{\delta }\bar{\psi}\vec{\tau}\psi , \\
\partial _{\nu }\vec{\rho}^{\mu \nu }+m_{\rho }^{2}\vec{\rho}^{\mu }
&=&\Gamma _{\rho }\bar{\psi}\vec{\tau}\gamma ^{\mu }\psi .
\end{eqnarray}

In the static case for an infinite nuclear matter, the meson
equations of motion become
\begin{eqnarray}
m_{\sigma }^{2}\bar{\sigma} &=&\Gamma _{\sigma }\rho _{S}, \\
m_{\omega }^{2}\bar{\omega}_{0} &=&\Gamma _{\omega }\rho _{B}, \\
m_{\rho }^{2}\bar{\rho}_{0}^{(3)} &=&\Gamma _{\rho }(\rho _{p}-\rho _{n}), \\
m_{\delta }^{2}{\bar{\delta}}^{(3)} &=&\Gamma _{\delta }(\rho _{S,p}-\rho
_{S,n}),
\end{eqnarray}
so the nucleon scalar and vector self-energies are
\begin{eqnarray}
\Sigma _{\tau }^{S} &=&-\Gamma _{\sigma }\bar{\sigma}-\Gamma _{\delta }{\bar{%
\delta}}^{(3)}\tau _{3}, \\
\Sigma _{\tau }^{0} &=&\Gamma _{\omega }\bar{\omega}_{0}+\Gamma _{\rho }\bar{
\rho}_{0}^{(3)}\tau _{3}+\Sigma ^{0(R)},
\end{eqnarray}
with
\begin{eqnarray}
\Sigma ^{0(R)} &=&\frac{\partial \Gamma _{\omega }}{\partial \rho }\bar{
\omega}_{0}\rho _{B}+\frac{\partial \Gamma _{\rho }}{\partial \rho }\bar{\rho%
}_{0}^{(3)}(\rho _{p}-\rho _{n})  \notag \\
&&-\frac{\partial \Gamma _{\sigma }}{\partial \rho }\bar{\sigma}\rho _{S}-%
\frac{\partial \Gamma _{\delta }}{\partial \rho }{\bar{\delta}}^{(3)}(\rho
_{S,p}-\rho _{S,n}).
\end{eqnarray}

\subsubsection{Nuclear matter equation of state}

From the energy-momentum tensor, the energy density and pressure of
nuclear matter can be derived, and they are given by
\begin{eqnarray}
\epsilon &=&\epsilon _{kin}^{n}+\epsilon _{kin}^{p}  \notag \\
&&+\frac{1}{2}\left[ m_{\sigma }^{2}\bar{\sigma}^{2}+m_{\omega }^{2}\bar{
\omega}_{0}^{2}+m_{\delta }^{2}{\bar{\delta}}^{(3)2}+m_{\rho }^{2}\bar{\rho}
_{0}^{(3)2}\right]
\end{eqnarray}
and
\begin{eqnarray}
P &=&P_{kin}^{n}+P_{kin}^{p}+\rho _{B}\Sigma ^{0(R)}  \notag \\
&&-\frac{1}{2}\left[ m_{\sigma }^{2}\bar{\sigma}^{2}-m_{\omega }^{2}\bar{
\omega}_{0}^{2}+m_{\delta }^{2}{\bar{\delta}}^{(3)2}-m_{\rho }^{2}\bar{\rho}
_{0}^{(3)2}\right] .
\end{eqnarray}
It is seen that the rearrangement self-energy does not affect the
energy density but contributes explicitly to the pressure.
Furthermore, the symmetry energy can be written as
\begin{eqnarray}
E_{\mathrm{sym}}(\rho _{B}) &=&\frac{k_{F}^{2}}{6\tilde{E}_{F}}+\frac{1}{2}
\left( \frac{\Gamma _{\rho }}{m_{\rho }}\right) ^{2}\rho _{B}-\frac{1}{2}
\left( \frac{\Gamma _{\delta }}{m_{\delta }}\right) ^{2}  \notag \\
&&\times \frac{M^{\ast 2}\rho _{B}}{\tilde{E}_{F}^{2}[1+\left( \frac{\Gamma
_{\delta }}{m_{\delta }}\right) ^{2}A(k_{F},M^{\ast })]},  \label{EsymDD}
\end{eqnarray}
with notations similarly defined as in the nonlinear RMF model.

\subsection{The nonlinear point-coupling RMF model}

\subsubsection{Lagrangian density}

The point-coupling model is defined by a Lagrangian density that
consists of only nucleon fields. In the present study, we use the
Lagrangian density of the nonlinear point-coupling model of
Refs.\cite{Nik92,Bur02}, i.e.,
\begin{equation}
\mathcal{L}_{\mathrm{NLPC}}=\mathcal{L}^{\mathrm{free}}+\mathcal{L}^{\mathrm{%
\ 4f}}+\mathcal{L}^{\mathrm{hot}}+\mathcal{L}^{\mathrm{der}},
\label{LagNLPC}
\end{equation}
with
\begin{eqnarray}
\mathcal{L}^{\mathrm{free}} &=&\bar{\psi}(\mathrm{i}\gamma _{\mu }\partial
^{\mu }-M)\psi , \\
\mathcal{L}^{\mathrm{4f}}\hfill &=&-{{\frac{1}{2}}}\,\alpha _{\mathrm{S}}(
\bar{\psi}\psi )(\bar{\psi}\psi )-{{\frac{1}{2}}}\,\alpha _{\mathrm{V}}(\bar{
\psi}\gamma _{\mu }\psi )(\bar{\psi}\gamma ^{\mu }\psi )  \notag \\
&&-{{\frac{1}{2}}}\,\alpha _{\mathrm{TS}}(\bar{\psi}\vec{\tau}\psi )\cdot (%
\bar{\psi}\vec{\tau}\psi )  \notag \\
&&-{{\frac{1}{2}}}\,\alpha _{\mathrm{TV}}(\bar{\psi}\vec{\tau}\gamma _{\mu
}\psi )\cdot (\bar{\psi}\vec{\tau}\gamma ^{\mu }\psi ), \\
\mathcal{L}^{\mathrm{hot}} &=&-{{\frac{1}{3}}}\,\beta _{\mathrm{S}}(\bar{\psi%
}\psi )^{3}-{{\frac{1}{4}}}\, \gamma _{\mathrm{S}}(\bar{\psi}\psi )^{4}
\notag \\
&&-{{\frac{1}{4}}}\,\gamma _{\mathrm{V}}[(\bar{\psi}\gamma _{\mu }\psi ) (%
\bar{\psi}\gamma ^{\mu }\psi )]^{2}  \notag \\
&&-{{\frac{1}{4}}}\,\gamma _{\mathrm{TV}}[(\bar{\psi}\vec{\tau}\gamma _{\mu
}\psi )\cdot (\bar{\psi}\vec{\tau}\gamma ^{\mu }\psi )]^{2}, \\
\mathcal{L}^{\mathrm{der}} &=&-{{\frac{1}{2}}}\,\delta _{\mathrm{S}
}(\partial _{\nu }\bar{\psi}\psi )(\partial ^{\nu }\bar{\psi}\psi )  \notag
\\
&&-{{\frac{1}{2}}}\,\delta _{\mathrm{V}}(\partial _{\nu }\bar{\psi}\gamma
_{\mu }\psi )(\partial ^{\nu }\bar{\psi}\gamma ^{\mu }\psi )  \notag \\
&&-{{\frac{1}{2}}}\,\delta _{\mathrm{TS}}(\partial _{\nu }\bar{\psi}\vec{\tau%
}\psi )\cdot (\partial ^{\nu }\bar{\psi}\vec{\tau}\psi )  \notag \\
&&-{{\frac{1}{2}}}\,\delta _{\mathrm{TV}}(\partial _{\nu }\bar{\psi}\vec{%
\tau }\gamma _{\mu }\psi )\cdot (\partial ^{\nu }\bar{\psi}\vec{\tau}\gamma
^{\mu }\psi ).
\end{eqnarray}
In the above, $\mathcal{L}^{\mathrm{free}}$ is the kinetic term of
nucleons and $\mathcal{L}^{ \mathrm{4f}}$ describes the
four-fermion interactions while $\mathcal{L}^{\mathrm{hot} } $ and
$\mathcal{L}^{\mathrm{der}}$ contain, respectively, higher-order
terms involving more than four fermions and derivatives in the
nucleon field. For the twelve coupling constants in the Lagrangian
density, $\alpha _{\mathrm{S}}$, $\alpha _{\mathrm{V}}$, $ \alpha
_{\mathrm{TS}}$, $\alpha _{\mathrm{TV}}$, $\beta _{\mathrm{S}}$, $
\gamma _{\mathrm{S}}$, $\gamma _{\mathrm{V}}$, $\gamma
_{\mathrm{TV}}$, $\delta _{\mathrm{S}}$, $\delta _{\mathrm{V}}$,
$\delta _{\mathrm{TS}}$, and $\delta _{\mathrm{TV}}$, the
subscripts denote the tensor structure of a coupling with
\textquotedblleft S\textquotedblright\, \textquotedblleft
V\textquotedblright\, and \textquotedblleft T\textquotedblright\
stand for scalar, vector, and isovector, respectively. The symbols
$\alpha _{\mathrm{i}}$, $\delta _{\mathrm{i}}$, $\beta _{\mathrm{i
}}$, and $\gamma _{\mathrm{i}} $ refer, respectively, to
four-fermion or second-order terms, derivative couplings, third-
and fourth order terms \cite{Nik92,Bur02}.

\subsubsection{Equation of motion and nucleon self-energies}

From the variation of the Lagrangian density Eq. (\ref{LagNLPC})
with respect to $\bar{\psi}$, we obtain the following Dirac equation
for the nucleon field:
\begin{equation}
\lbrack \gamma _{\mu }(i\partial ^{\mu }-\Sigma ^{\mu })-(M+\Sigma
^{S})]\psi =0,
\end{equation}
where the nucleon scalar ($\Sigma ^{S}$) and vector ($\Sigma ^{\mu
}$) self-energies are
\begin{eqnarray}
\Sigma ^{S}&=&V_{S}+\vec{V}_{TS}\cdot \vec{\tau},\\
\Sigma ^{\mu }&=&V^{\mu }+\vec{V}_{T}^{\mu }\cdot \vec{\tau},
\end{eqnarray}
respectively, with
\begin{eqnarray}
V_{S} &=&\alpha _{\mathrm{S}}(\bar{\psi}\psi )+\beta _{\mathrm{S}}(\bar{\psi}
\psi )^{2}+\gamma _{\mathrm{S}}(\bar{\psi}\psi )^{3}  \notag \\
&&-\delta _{\mathrm{S}}\square (\bar{\psi}\psi ), \\
\vec{V}_{TS} &=&\alpha _{\mathrm{TS}}(\bar{\psi}\vec{\tau}\psi )  \notag \\
&&-\delta _{\mathrm{TS}}\square (\bar{\psi}\vec{\tau}\psi ), \\
V^{\mu } &=&\alpha _{\mathrm{V}}(\bar{\psi}\gamma ^{\mu }\psi )+\gamma _{%
\mathrm{V}}(\bar{\psi}\gamma ^{\mu }\psi ) (\bar{\psi}\gamma _{\mu }\psi )(%
\bar{\psi}\gamma ^{\mu }\psi )  \notag \\
&&-\delta _{\mathrm{V}}\square (\bar{\psi}\gamma ^{\mu }\psi ), \\
\vec{V}_{T}^{\mu } &=&\alpha _{\mathrm{TV}}(\bar{\psi}\vec{\tau}\gamma ^{\mu
}\psi )+\gamma _{\mathrm{TV}}(\bar{\psi}\vec{\tau}\gamma ^{\mu }\psi )\cdot
( \bar{\psi}\vec{\tau}\gamma _{\mu }\psi )(\bar{\psi}\vec{\tau}\gamma
^{\mu}\psi )  \notag \\
&&-\delta _{\mathrm{TV}}\square (\bar{\psi}\vec{\tau}\gamma ^{\mu }\psi ).
\end{eqnarray}
In the above, $\square =\partial ^{2}/(c^{2}\partial
t^{2}-\bigtriangleup )$ denotes the four-dimensional
d'Alembertian. In the translationally invariant infinite nuclear
matter, all terms involving derivative couplings drop out and the
spatial components of the four-currents also vanish. In terms of
the baryon density $\rho_B$ and scalar density $\rho_S$ as well as
the isospin baryon density $\rho_3=\rho_p-\rho_n$ and the isospin
scalar density $\rho_{S3}=\rho_{S,p}-\rho_{S,n}$, the nucleon
scalar and vector self-energies in asymmetric nuclear matter can
be rewritten as
\begin{eqnarray}
\Sigma _{\tau }^{S} &=&\alpha _{\mathrm{S}}\rho _{S}+\beta _{\mathrm{S}}\rho
_{S}^{2}+\gamma _{\mathrm{S}}\rho _{S}^{3}+\alpha _{\mathrm{TS}}\rho
_{S3}\tau _{3}, \\
\Sigma _{\tau }^{0} &=&\alpha _{\mathrm{V}}\rho _{B}+\gamma _{\mathrm{V}
}\rho _{B}^{3}+\alpha _{\mathrm{TV}}\rho _{3}\tau _{3}+\gamma _{\mathrm{TV}%
}\rho _{3}^{3}\tau _{3}.  \label{Sig0NLPC}
\end{eqnarray}

\subsubsection{Nuclear matter equation of state}

The energy density $\epsilon $ and the pressure $P$ derived from
the energy-momentum tensor in the nonlinear point-coupling RMF
model are given by
\begin{eqnarray}
\epsilon &=&\epsilon _{kin}^{n}+\epsilon _{kin}^{p}-\frac{1}{2}\alpha _{
\mathrm{S}}\rho _{S}^{2}-\frac{1}{2}\alpha _{\mathrm{TS}}\rho _{S3}^{2}
\notag \\
&&+\frac{1}{2}\alpha _{\mathrm{V}}\rho ^{2}+\frac{1}{2}\alpha _{\mathrm{TV}%
}\rho _{3}^{2}\;  \notag \\
&&-\frac{1}{3}\beta _{\mathrm{S}}\rho _{S}^{3}-\frac{3}{4}\gamma _{\mathrm{S}
}\rho _{S}^{4}+\frac{1}{4}\gamma _{\mathrm{V}}\rho ^{4}+\frac{1}{4}\gamma _{
\mathrm{TV}}\rho _{3}^{4},
\end{eqnarray}
\begin{eqnarray}
P &=&\tilde{E}_{F}^{p}\rho _{p}+\tilde{E}_{F}^{n}\rho _{n}-\epsilon
_{kin}^{p}-\epsilon _{kin}^{n}  \notag \\
&&+\frac{1}{2}\alpha _{\mathrm{S}}\rho _{s}^{2}+\frac{1}{2}\alpha _{\mathrm{%
\ TS}}\rho _{s3}^{2}+\frac{1}{2}\alpha _{\mathrm{V}}\rho ^{2}+\frac{1}{2}
\alpha _{\mathrm{TV}}\rho _{3}^{2}  \notag \\
&&+\frac{2}{3}\beta _{\mathrm{S}}\rho _{s}^{3}+\frac{3}{4}\gamma
_{\mathrm{S} }\rho _{s}^{4}+\frac{3}{4}\gamma _{\mathrm{V}}\rho
^{4}+\frac{3}{4}\gamma _{ \mathrm{TV}}\rho _{3}^{4},
\end{eqnarray}
where $\tilde{E}_{F}^{p}$ and $\tilde{E}_{F}^{n}$ are defined as in Eq. (\ref%
{Ef}) with the nucleon Dirac masses
\begin{eqnarray}
M_{p}^{\ast } &=&\alpha _{\mathrm{S}}\rho _{S}+\beta _{\mathrm{S}}\rho
_{S}^{2}+\gamma _{\mathrm{S}}\rho _{S}^{3}+\alpha _{\mathrm{TS}}\rho _{S3},
\\
M_{n}^{\ast } &=&\alpha _{\mathrm{S}}\rho _{S}+\beta _{\mathrm{S}}\rho
_{S}^{2}+\gamma _{\mathrm{S}}\rho _{S}^{3}-\alpha _{\mathrm{TS}}\rho _{S3}.
\end{eqnarray}
Furthermore, the symmetry energy in this model can be expressed as
\begin{eqnarray}
E_{\mathrm{sym}}(\rho _{B}) &=&\frac{k_{F}^{2}}{6\tilde{E}_{F}}+\frac{1}{2}
\alpha _{\mathrm{TV}}\rho _{B}  \notag \\
&&+\frac{1}{2}\alpha _{\mathrm{TS}}\frac{M^{\ast 2}\rho _{B}}{\tilde{E}
_{F}^{2}[1-\alpha _{\mathrm{TS}}A(k_{F},M^{\ast })]},  \label{EsymNLPC}
\end{eqnarray}
with notations again similarly defined as in the nonlinear RMF model.

\subsection{The density-dependent point-coupling RMF model}

\subsubsection{Lagrangian density}

For the density-dependent point-coupling RMF model, we use the
Lagrangian density of Refs.\cite{Fin04,Fin06}, i.e.,
\begin{equation}
\mathcal{L}_{\text{DDPC}}=\mathcal{L}_{\mathrm{free}}+\mathcal{L}_{\mathrm{%
4f }}+\mathcal{L}_{\mathrm{der}},  \label{LagDDPC}
\end{equation}
with
\begin{eqnarray}
\mathcal{L}_{\mathrm{free}} &=&\bar{\psi}(i\gamma _{\mu }\partial ^{\mu
}-M)\psi , \\
\mathcal{L}_{\mathrm{4f}} &=&-\frac{1}{2}~G_{S}(\hat{\rho})(\bar{\psi}\psi
)( \bar{\psi}\psi )  \notag \\
&~&-\frac{1}{2}~G_{V}(\hat{\rho})(\bar{\psi}\gamma _{\mu }\psi )(\bar{\psi}%
\gamma ^{\mu }\psi )  \notag \\
&~&-\frac{1}{2}~G_{TS}(\hat{\rho})(\bar{\psi}\vec{\tau}\psi )\cdot (\bar{\psi%
}\vec{\tau}\psi )  \notag \\
&~&-\frac{1}{2}~G_{TV}(\hat{\rho})(\bar{\psi}\vec{\tau}\gamma _{\mu }\psi
)\cdot (\bar{\psi}\vec{\tau}\gamma ^{\mu }\psi ), \\
\mathcal{L}_{\mathrm{der}} &=&-\frac{1}{2}~D_{S}(\hat{\rho})(\partial _{\nu
} \bar{\psi}\psi )(\partial ^{\nu }\bar{\psi}\psi ).
\end{eqnarray}
In the above, $\mathcal{L}^{\mathrm{free}}$ is the kinetic term of
the nucleons and $\mathcal{L}^{\mathrm{4f}}$ is a four-fermion
interaction while $\mathcal{L}^{\mathrm{der}}$ represents
derivatives in the nucleon scalar densities. Unlike in the nonlinear
point-coupling RMF model, the density-dependent point-coupling RMF
model used here includes only second-order interaction terms with
density-dependent couplings $G_{i}(\hat{\rho})$ and
$D_{i}(\hat{\rho}) $ that are determined from finite-density QCD sum
rules and in-medium chiral perturbation theory \cite{Fin04,Fin06}.

\subsubsection{Equation of motion and nucleon self-energies}

Variation of the Lagrangian Eq.(\ref{LagDDPC}) with respect to $\bar{\psi}$
leads to the single-nucleon Dirac equation
\begin{equation}
\lbrack \gamma _{\mu }(i\partial ^{\mu }-\Sigma ^{\mu })-(M+\Sigma
^{S})]\psi =0,
\end{equation}
with the nucleon scalar and vector self-energies given, respectively, by
\begin{eqnarray}
\Sigma ^{S}&=&V_{S}+\vec{V}_{TS}\cdot \vec{\tau}+\Sigma _{rS}\;,\\
\Sigma ^{\mu }&=&V^{\mu }+\vec{V}_{T}^{\mu }\cdot \vec{\tau}+\Sigma
_{r}^{\mu },
\end{eqnarray}
where
\begin{eqnarray}
V_{S} &=&G_{S}(\bar{\psi}\psi )-D_{S}\square (\bar{\psi}\psi ), \\
\vec{V}_{TS} &=&G_{TS}(\bar{\psi}\vec{\tau}\psi ), \\
V^{\mu } &=&G_{V}(\bar{\psi}\gamma ^{\mu }\psi ), \\
\vec{V}_{T}^{\mu } &=&G_{TV}(\bar{\psi}\vec{\tau}\gamma ^{\mu }\psi ), \\
\Sigma _{rS} &=&-\frac{\partial D_{S}}{\partial \hat{\rho}}(\partial _{\nu
}j^{\mu })u_{\mu }(\partial ^{\nu }(\bar{\psi}\psi ))
\end{eqnarray}
and
\begin{eqnarray}
\Sigma _{r}^{\mu } &=&\frac{u^{\mu }}{2}\left( \frac{\partial G_{S}}{%
\partial \hat{\rho}}(\bar{\psi}\psi )(\bar{\psi}\psi )+\frac{\partial G_{TS}
}{\partial \hat{\rho}}(\bar{\psi}\vec{\tau}\psi )\cdot (\bar{\psi}\vec{\tau}
\psi )\right.  \notag \\
&~&+\frac{\partial G_{V}}{\partial \hat{\rho}}(\bar{\psi}\gamma ^{\mu }\psi
)(\bar{\psi}\gamma _{\mu }\psi )+\frac{\partial G_{TV}}{\partial \hat{\rho}}%
( \bar{\psi}\vec{\tau}\gamma ^{\mu }\psi )\cdot (\bar{\psi}\vec{\tau}\gamma
_{\mu }\psi )  \notag \\
&~&\left. +\frac{\partial D_{S}}{\partial \hat{\rho}}(\partial ^{\nu }(\bar{
\psi}\psi ))(\partial _{\nu }(\bar{\psi}\psi ))\right).
\end{eqnarray}
In the above, we have $\hat{\rho}u^{\mu }=\bar{\psi}\gamma ^{\mu
}\psi $, where the four-velocity $u^{\mu }$ is defined as
$(1-\mathbf{v}^{2})^{-1/2}(1,\mathbf{v})$ with $\mathbf{v}$ being
the three-velocity vector, and $\Sigma _{rS}$ and $\Sigma
_{r}^{\mu } $ represent the rearrangement contributions resulting
from the variation of the vertex functionals with respect to the
nucleon fields in the density operator $\hat{\rho}$. The latter
coincides with the baryon density in the nuclear matter rest
frame.

In the translationally invariant infinite asymmetric nuclear matter,
the nucleon scalar and vector self-energies become
\begin{eqnarray}
\Sigma _{\tau }^{S} &=&G_{S}\rho _{S}+G_{TS}\rho _{S3}\tau _{3} \\
\Sigma _{\tau }^{0} &=&G_{V}\rho _{B}+G_{TV}\rho _{3}\tau _{3}+\Sigma
^{0(R)},
\end{eqnarray}
with the rearrangement contribution to the self-energy
\begin{equation}
\Sigma ^{0(R)}=\frac{1}{2}[\frac{\partial G_{S}}{\partial \rho }\rho
_{S}^{2}+\frac{\partial G_{TS}}{\partial \rho }\rho _{S3}^{2}+\frac{\partial
G_{V}}{\partial \rho }\rho ^{2}+\frac{\partial G_{TV}}{\partial \rho }\rho
_{3}^{2}].
\end{equation}

\subsubsection{Nuclear matter equation of state}

For asymmetric nuclear matter, the energy density $\epsilon $ and
the pressure $P$ derived from the energy-momentum tensor in the
density-dependent point-coupling RMF model are
\begin{eqnarray}
\epsilon &=&\epsilon _{kin}^{n}+\epsilon _{kin}^{p}-\frac{1}{2}G_{S}\rho
_{S}^{2}-\frac{1}{2}G_{TS}\rho _{S3}^{2}  \notag \\
&&+\frac{1}{2}G_{V}\rho ^{2}+\frac{1}{2}G_{TV}\rho _{3}^{2},
\end{eqnarray}
and
\begin{eqnarray}
P &=&\tilde{E}_{F}^{p}\rho _{p}+\tilde{E}_{F}^{n}\rho _{n}-\epsilon
_{kin}^{p}-\epsilon _{kin}^{n}  \notag \\
&&+\frac{1}{2}G_{V}\rho ^{2}+\frac{1}{2}G_{TV}\rho _{3}^{2}+\frac{1}{2}
G_{S}\rho _{S}^{2}+\frac{1}{2}G_{TS}\rho _{S3}^{2}  \notag \\
&~&+\frac{1}{2}\frac{\partial G_{S}}{\partial \rho }\rho _{S}^{2}\rho +\frac{
1}{2}\frac{\partial G_{V}}{\partial \rho }\rho ^{3}  \notag \\
&&+\frac{1}{2}\frac{\partial G_{TV}}{\partial \rho }\rho _{3}^{2}\rho +\frac{
1}{2}\frac{\partial G_{TS}}{\partial \rho }\rho _{S3}^{2}\rho,
\end{eqnarray}
where $\tilde{E}_{F}^{p}$ and $\tilde{E}_{F}^{n}$ are defined as in
Eq. (\ref{Ef}) with the effective nucleon masses
\begin{eqnarray}
M_{p}^{\ast } &=&M+G_{S}\rho _{S}+G_{TS}\rho _{S3}, \\
M_{n}^{\ast } &=&M+G_{S}\rho _{S}-G_{TS}\rho _{S3}.
\end{eqnarray}
As in the density-dependent RMF model, \textit{rearrangement}
contributions appear explicitly in the expression for the
pressure. Finally, the symmetry energy can be written as
\begin{eqnarray}
E_{\mathrm{sym}}(\rho _{B}) &=&\frac{k_{F}^{2}}{6\tilde{E}_{F}}+\frac{1}{2}
G_{TV}\rho _{B}  \notag \\
&&+\frac{1}{2}G_{TS}\frac{M^{\ast 2}\rho _{B}}{\tilde{E}
_{F}^{2}[1-G_{TS}A(k_{F},M^{\ast })]},  \label{EsymDDPC}
\end{eqnarray}
with similar notations as in the nonlinear RMF model.

\section{Results and discussions}

\label{results}

Using above models, we have studied the isospin-dependent
properties of asymmetric nuclear matter. In the following, we
focus on results regarding the nuclear symmetry energy, the
nuclear symmetry potential, the isospin-splitting of nucleon
effective mass, and the isospin-dependent nucleon scalar density
in asymmetric nuclear matter. For different versions of the RMF
model considered in the present work, we mainly consider parameter
sets commonly and successfully used in nuclear structure studies.
In particular, we select the parameter sets NL1 \cite{Lee86}, NL2
\cite{Lee86}, NL3 \cite{Lal97}, NL-SH \cite{Sha93}, TM1
\cite{Sug94}, PK1 \cite{Lon04}, FSU-Gold \cite{Tod05}, HA
\cite{Bun03}, NL$\rho $ \cite{Liu02}, NL$\rho \delta $
\cite{Liu02} for the nonlinear RMF model; TW99 \cite{Typ99},
DD-ME1 \cite{Nik02}, DD-ME2 \cite{Lal05}, PKDD \cite{Lon04}, DD
\cite{Typ05}, DD-F \cite{Kla06}, and DDRH-corr \cite{Hof01} for
the density-dependent RMF model; and PC-F1 \cite{Bur02}, PC-F2
\cite{Bur02}, PC-F3 \cite{Bur02}, PC-F4 \cite{Bur02}, PC-LA
\cite{Bur02}, and FKVW \cite{Fin06} for the point-coupling RMF
model. There are totally $23$ parameter sets, and most of them can
describe reasonably well the binding energies and charge radii of
a large number of nuclei in the periodic table except the
parameter set HA, for which to our knowledge there are no
calculations for finite nuclei.

We note that all selected parameter sets include the
isovector-vector channel involving either the isovector-vector
$\rho $ meson or the isovector-vector interaction vertices in the
Lagrangian. The HA parameter set further includes the
isovector-scalar meson field $\vec{\delta}$ and fits successfully
some results obtained from the more microscopic DBHF approach
\cite{Bun03}. The parameter sets NL$\rho \delta $ and DDRH-corr
also include the isovector-scalar meson field $\vec{\delta}$,
while PC-F2, PC-F4, PC-LA, and FKVW include the isovector-scalar
interaction vertices. The parameter sets NL$\rho \delta $ as well
as NL$\rho $ are obtained from fitting the empirical properties of
asymmetric nuclear matter \cite{Liu02} and describe reasonably
well the binding energies and charge radii of a large number of
nuclei \cite{Gai04}. For the DDRH-corr, its parameters are
determined from the density-dependent meson-nucleon vertices
extracted from the self-energies of asymmetric nuclear matter
calculated in the microscopic DBHF approach with momentum
corrections, and it reproduces satisfactorily the properties of
finite nuclei and the EOS from the DBHF approach \cite{Hof01}. In
the parameter sets PC-F1, PC-F2, PC-F3, PC-F4 and PC-LA for the
nonlinear point-coupling model, their coupling constants are
determined in a self-consistent procedure that solves the model
equations for representative nuclei simultaneously in a
generalized nonlinear least-squares adjustment algorithm
\cite{Bur02}. The parameter set FKVW for the density-dependent
point-coupling model are determined by the constraints derived
from the finite-density QCD sum rules, in-medium chiral
perturbation theory, and experimental data of a number of finite
nuclei \cite{Fin06}.

\subsection{Nuclear symmetry energy}

\begin{figure}[th]
\includegraphics[scale=1.2]{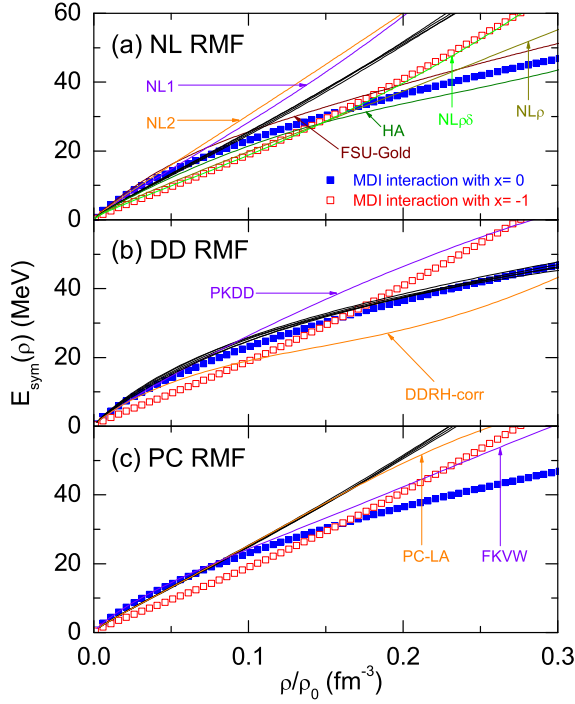}
\caption{{\protect\small (Color online) Density dependence of the
nuclear symmetry energy }$E_{\mathrm{sym}}(\protect\rho
)${\protect\small \ for the parameter sets NL1, NL2, NL3, NL-SH,
TM1, PK1, FSU-Gold, HA, NL}$\protect \rho ${\protect\small , and
NL}$\protect\rho \protect\delta ${\protect\small \ in the
nonlinear RMF model (a); TW99, DD-ME1, DD-ME2, PKDD, DD, DD-F, and
DDRH-corr in the density-dependent RMF model (b); and PC-F1,
PC-F2, PC-F3, PC-F4, PC-LA, and FKVW in the point-coupling RMF
model (c). For comparison, results from the MDI interaction with
}$x=-1${\protect\small \ (open squares) and }$0${\protect\small \
(solid squares) are also shown.}} \label{EsymDen}
\end{figure}

Fig. \ref{EsymDen} displays the density dependence of the nuclear
symmetry energy $E_{\mathrm{sym}}(\rho )$ for the $23$ parameter
sets in the nonlinear, density-dependent, and point-coupling RMF
models. For comparison, we also show in Fig. \ref{EsymDen} results
from the phenomenological parametrization of the
momentum-dependent nuclear mean-field potential based on the Gogny
effective interaction \cite{Das03}, i.e., the MDI interactions
with $x=-1$ (open squares) and $0$ (solid squares), where
different $x$ values correspond to different density dependence of
the nuclear symmetry energy but keep other properties of the
nuclear EOS the same \cite{Che05a} (see Appendix \ref{MDI} for
details). From analyzing the isospin diffusion data from NSCL/MSU
using the IBUU04 transport model with in-medium NN cross sections,
it has been found that the MDI interactions with $x=-1$ and $0$
give, respectively, the upper and lower bounds for the stiffness
of the nuclear symmetry energy at densities up to about
$1.2\rho_0$ \cite{Che05a,LiBA05c}.

It is seen from Fig. \ref{EsymDen} that the density dependence of
symmetry energy varies drastically among different interactions.
In the nonlinear RMF model, while the dependence on density is
almost linear for most parameter sets, it is much softer for the
parameter sets FSU-Gold and HA. The softening of the symmetry
energy from the latter two parameter sets is due to the mixed
isoscalar-isovector couplings $\Lambda _{S}$ and $\Lambda _{V}$
\cite{Mul96,Hor01a} which modifies the density dependence of
symmetry energy as seen in Eq. (\ref{EsymNL}). For the parameter
set NL$\rho \delta $, it gives a symmetry energy that depends
linearly on density at low densities but becomes stiffer at high
densities due to inclusion of the isovector-scalar $\delta $
meson. The approximate linear density-dependent behavior of the
symmetry energy for other parameter sets in the nonlinear RMF
model can also be understood from Eq. (\ref{EsymNL}), which shows
that the symmetry energy at high densities is dominated by the
potential energy that is proportional to the baryon density if the
mixed isoscalar-isovector coupling and the isovector-scalar
$\delta $ meson are not included in the model.

The density dependence of the symmetry energy in the
density-dependent RMF model is essentially determined by the
density dependence of the coupling constants $\Gamma _{\rho }$ and
$\Gamma _{\delta }$ of isovector mesons. Most parameter sets in
this case give similar symmetry energies except the parameter sets
PKDD and DDRH-corr. Compared with other parameter sets in the
density-dependent RMF model, PKDD gives a very large while
DDRH-corr gives a very small value for the symmetry energy at
saturation density. For point-coupling models, all parameter sets
(PC-F1, PC-F2, PC-F3, PC-F4 and PC-LA) in the nonlinear
point-coupling RMF model predict almost linearly density-dependent
symmetry energies while the parameter set FKVW in the
density-dependent point-coupling RMF model gives a somewhat softer
symmetry energy.

\begin{table}[tbp]
\caption{{\protect\small Bulk properties of nuclear matter at the
saturation point: }$-B/A${\protect\small \ (MeV), }$\protect\rho
_{0}${\protect\small \ (fm}$^{-3}${\protect\small ),
}$K_{0}${\protect\small \ (MeV), }$E_{\text{sym }}(\protect\rho
_{0})${\protect\small \ (MeV), }$K_{\text{sym}}$ {\protect\small \
(MeV), }$L${\protect\small \ (MeV), and }$K_{\text{asy}}$
{\protect\small \ (MeV) using the }$23${\protect\small \ parameter
sets in the nonlinear, density-dependent, and point-coupling RMF
models. The last column gives the references for corresponding
parameter sets.}}
\label{Bulk}%
\begin{tabular}{ccccccccc}
\hline\hline
Model & $\quad -B/A$ & $\rho _{0}$ & $K_{0}$ & $E_{\text{sym}}$ & $L$ & $K_{%
\text{sym}}$ & $K_{\text{asy}}$ & Ref. \\ \hline
NL1 & $16.4$ & $0.152$ & $212$ & $43.5$ & $140$ & $143$ & $-697$ & \cite%
{Lee86} \\
NL2 & $17.0$ & $0.146$ & $401$ & $44.0$ & $130$ & $20$ & $-750$ & \cite%
{Lee86} \\
NL3 & $16.2$ & $0.148$ & $271$ & $37.3$ & $118$ & $100$ & $-608$ & \cite%
{Lal97} \\
NL-SH & $16.3$ & $0.146$ & $356$ & $36.1$ & $114$ & $80$ & $-604$ & \cite%
{Sha93} \\
TM1 & $16.3$ & $0.145$ & $281$ & $36.8$ & $111$ & $34$ & $-632$ & \cite%
{Sug94} \\
PK1 & $16.3$ & $0.148$ & $282$ & $37.6$ & $116$ & $55$ & $-641$ & \cite%
{Lon04} \\
FSUGold & $16.3$ & $0.148$ & $229$ & $32.5$ & $60$ & $-52$ & $-412$ & \cite%
{Tod05} \\
HA & $15.6$ & $0.170$ & $233$ & $30.7$ & $55$ & $-135$ & $-465$ & \cite%
{Bun03} \\
NL$\rho $ & $16.1$ & $0.160$ & $240$ & $30.3$ & $85$ & $3$ & $-507$ & \cite%
{Liu02} \\
NL$\rho \delta $ & $16.1$ & $0.160$ & $240$ & $30.7$ & $103$ & $127$ & $-491$
& \cite{Liu02} \\
&  &  &  &  &  &  &  &  \\
TW99 & $16.2$ & $0.153$ & $241$ & $32.8$ & $55$ & $-124$ & $-454$ & \cite%
{Typ99} \\
DD-ME1 & $16.2$ & $0.152$ & $245$ & $33.1$ & $55$ & $-101$ & $-431$ & \cite%
{Nik02} \\
DD-ME2 & $16.1$ & $0.152$ & $251$ & $32.3$ & $51$ & $-87$ & $-393$ & \cite%
{Lal05} \\
PKDD & $16.3$ & $0.150$ & $263$ & $36.9$ & $90$ & $-80$ & $-620$ & \cite%
{Lon04} \\
DD & $16.0$ & $0.149$ & $241$ & $31.7$ & $56$ & $-95$ & $-431$ & \cite{Typ05}
\\
DD-F & $16.0$ & $0.147$ & $223$ & $31.6$ & $56$ & $-140$ & $-476$ & \cite%
{Kla06} \\
DDRH-corr & $15.6$ & $0.180$ & $281$ & $26.1$ & $51$ & $155$ & $-151$ & \cite%
{Hof01} \\
&  &  &  &  &  &  &  &  \\
PC-F1 & $16.2$ & $0.151$ & $255$ & $37.8$ & $117$ & $75$ & $-627$ & \cite%
{Bur02} \\
PC-F2 & $16.2$ & $0.151$ & $256$ & $37.6$ & $116$ & $65$ & $-631$ & \cite%
{Bur02} \\
PC-F3 & $16.2$ & $0.151$ & $256$ & $38.3$ & $119$ & $74$ & $-640$ & \cite%
{Bur02} \\
PC-F4 & $16.2$ & $0.151$ & $255$ & $37.7$ & $119$ & $98$ & $-616$ & \cite%
{Bur02} \\
PC-LA & $16.1$ & $0.148$ & $263$ & $37.2$ & $108$ & $-61$ & $-709$ & \cite%
{Bur02} \\
FKVW & $16.2$ & $0.149$ & $379$ & $33.1$ & $80$ & $11$ & $-469$ & \cite%
{Fin06} \\ \hline\hline
\end{tabular}%
\end{table}

Fig. \ref{EsymDen} thus shows that only a few parameter sets can
give symmetry energies that are consistent with the constraint
from the isospin diffusion data in heavy-ion collisions, which is
given by results from the MDI interactions with $x=-1$ and $0$.
The main reason for this is that most parameter sets in the RMF
model have saturation densities and symmetry energies at their
saturation densities which are significantly different from the
empirical saturation density of $0.16$ fm$^{-3}$ and symmetry
energy of $31.6$ MeV at this saturation density. To show this more
clearly, we list in Table \ref{Bulk} the bulk properties of
nuclear matter at saturation density: the binding energy per
nucleon $-B/A$\ (MeV), the saturation density of symmetric nuclear
matter $\rho _{0}$\ (fm$^{-3}$), the incompressibility of
symmetric nuclear matter $K_{0}$\ (MeV), the symmetry energy
$E_{\mathrm{sym} }(\rho _{0})$\ (MeV), $K_{\mathrm{sym}}$\ (MeV),
$L$\ (MeV), and $K_{\mathrm{\ asy}}$\ (MeV) using the $23$
parameter sets in the nonlinear, density-dependent, and
point-coupling RMF models. It is seen that these parameter sets
give saturation densities varying from $\rho _{0}=0.145$ fm$^{-3}$
to $\rho _{0}=0.180$ fm$^{-3}$ and nuclear symmetry energies
$E_{\mathrm{sym}}(\rho _{0})$\ (MeV) ranging from $26.1$ to $44.0$
MeV.

\begin{figure}[th]
\includegraphics[scale=1.1]{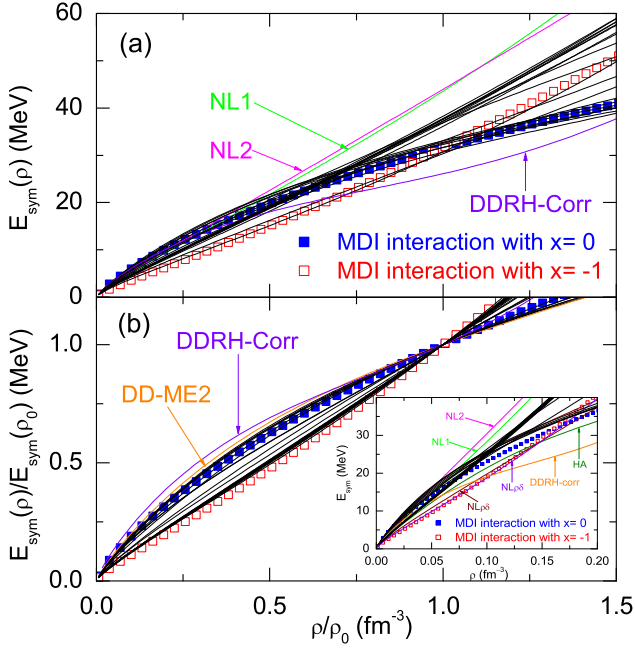}
\caption{{\protect\small (Color online) The symmetry energy }$E_{\mathrm{sym}
}(\protect\rho )${\protect\small \ (a) and the scaled symmetry energy }$E_{
\mathrm{sym}}(\protect\rho )/E_{\mathrm{sym}}(\protect\rho _{0})$
{\protect\small \ (b) as functions of the scaled baryon density }$\protect%
\rho /\protect\rho _{0}${\protect\small \ for the 23 parameter sets in the
nonlinear, density-dependent, and point-coupling RMF models. Results of the
MDI interaction with }$x=-1${\protect\small \ (open squares) and }$0$
{\protect\small \ (solid squares) are also included for comparison. The
inset in panel (b) shows the symmetry energy }$E_{\mathrm{sym}}(\protect%
\rho )${\protect\small \ as a function of the baryon density }$\protect\rho $
{\protect\small \ without scaling.}}
\label{EsymDenSCL}
\end{figure}

\begin{figure*}[th]
\includegraphics[scale=1.3]{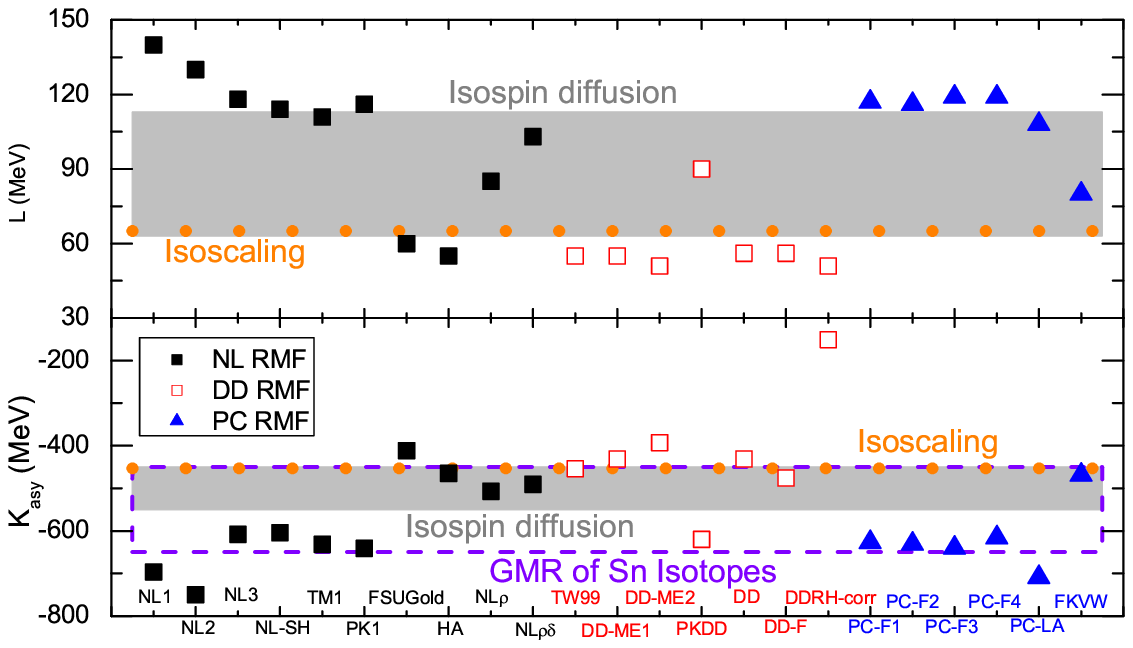}
\caption{{\protect\small (Color online) Values of
}$L${\protect\small \ and } $K_{\mathrm{asy}}${\protect\small \
for the }$23${\protect\small \ parameter sets in the nonlinear
(solid squares), density-dependent (open squares), and
point-coupling (triangles) RMF models. The constraints from the
isospin diffusion data (shaded band), the isoscaling data (solid
circles), and the isotopic dependence of the GMR in even-A Sn
isotopes (dashed rectangle) are also included.}} \label{LKasy}
\end{figure*}

To remove the effect due to differences in the saturation
densities among different parameter sets, we show in Fig.
\ref{EsymDenSCL} both the symmetry energy $E_{\mathrm{sym}}(\rho
)$ and the symmetry energy scaled by its value at corresponding
saturation density, i.e., $ E_{\mathrm{sym}}(\rho
)/E_{\mathrm{sym}}(\rho _{0})$ as functions of the scaled baryon
density $\rho /\rho _{0}$ for different parameter sets. For
comparison, we also plot in the inset in panel (b) of Fig.
\ref{EsymDenSCL} the symmetry energy $E_{\mathrm{sym}}(\rho )$ as
a function of the baryon density $\rho $ without scaling. It is
seen that more parameter sets among the $23$ sets become
consistent with the constraint from the isospin diffusion data in
heavy-ion collisions after scaling the baryon density by the
saturation density, and with further scaling of the symmetry
energy by its value at corresponding saturation density, most of
the parameter sets are in agreement with the constraint from the
isospin diffusion data. It is also interesting to see from the
inset in Fig. \ref{EsymDenSCL} that most of the parameter sets
obtained from fitting the properties of finite nuclei give roughly
the same value of about $26$ MeV for the nuclear symmetry energy
at the same baryon density of $\rho =0.1$ fm$^{-3}$. This
interesting feature is very similar to that found with Skyrme
interactions \cite{Bro00,Che05b}. It implies that the constraint
on the symmetry energy from fitting the properties of finite
nuclei is particularly sensitive to the nuclear properties at
lower densities, i.e., at density slightly above half-saturation
density.

For the density dependence of the nuclear symmetry energy around
saturation density, a more reasonable and physically meaningful
comparison is through the values of $L$ and $K_{\mathrm{asy}}$
given by these parameter sets since the $L$ parameter is
correlated linearly to the neutron-skin thickness of finite nuclei
while the $K_{\mathrm{asy}}$ parameter determines the isotopic
dependence of the GMR for a fixed element. From Table \ref{Bulk},
we have seen that the values of $L$, $K_{\mathrm{sym}}$, and
$K_{\mathrm{asy}}$ vary drastically, and they are in the range of
$51\sim 140$ MeV, $-140\sim 143$ MeV and $-750\sim $ $-151$ MeV,
respectively. The extracted values of $L=88\pm 25$ MeV and $K_{
\mathrm{asy}}=-500\pm 50$ MeV from the isospin diffusion data,
$L\approx 65$ MeV and $K_{\mathrm{asy}}\approx -453$ MeV from the
isoscaling data, and $K_{ \mathrm{asy}}=-550\pm 100$ MeV from the
isotopic dependence of the GMR in even-A Sn isotopes give a rather
stringent constraint on the density dependence of the nuclear
symmetry energy and thus put strong constraints on the nuclear
effective interactions as well. To see this constraint more
clearly, we collect in Fig. \ref{LKasy} the values of $L$ and
$K_{\mathrm{asy }}$ obtained from the $23$ parameter sets in the
nonlinear, density-dependent, and point-coupling RMF models
together with the constraints from the isospin diffusion data,
isoscaling data, and the isotopic dependence of the GMR in even-A
Sn isotopes. From Fig. \ref{LKasy} as well as Table \ref{Bulk}, we
see clearly that among the $23$ parameter sets considered here,
only six sets, i.e., TM1, NL$\rho $, NL$\rho \delta $, PKDD,
PC-LA, and FKVW, have nuclear symmetry energies that are
consistent with the extracted $L$ value of $88\pm 25$ MeV while
fifteen sets, i.e., NL3, NL-SH, TM1, PK1, HA, NL$\rho $, NL$\rho
\delta $, TW99, PKDD, DD-F, PC-F1, PC-F2, PC-F3, PC-F4, and FKVW,
have nuclear symmetry energies that are consistent with the
extracted $K_{\mathrm{asy}}$ value of $-500\pm 50$ MeV or $-550\pm
100$ MeV. Among the latter fifteen sets, only six sets, i.e., HA,
NL$\rho $, NL$\rho \delta $, TW99, DD-F, and FKVW are consistent
with $K_{\mathrm{asy}}=-500\pm 50$ MeV. It is interesting to see
that most parameter sets in the nonlinear and point-coupling RMF
models predict stiffer symmetry energies (i.e., larger values for
the $L$ parameter and larger magnitudes for $K_{\mathrm{asy}} $)
while those in the density-dependent RMF model give softer
symmetry energies (i.e., smaller values for the $L$ parameter and
smaller magnitudes for $K_{\mathrm{asy}}$).

\begin{figure}[th]
\includegraphics[scale=1.0]{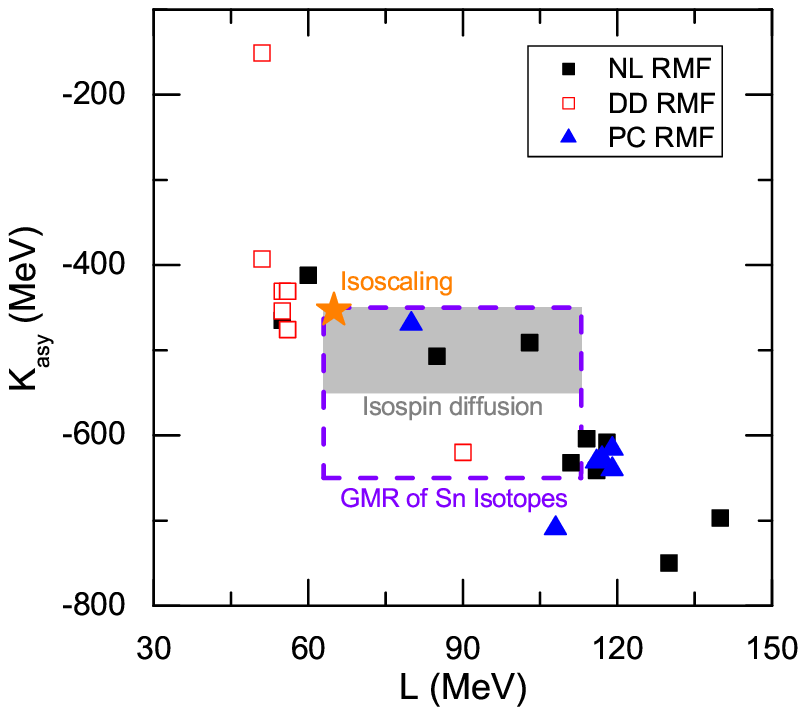}
\caption{{\protect\small (Color online) Correlation between }$L$
{\protect\small \ and }$K_{\text{asy}}${\protect\small \ for the
23 parameter sets in the nonlinear (solid squares),
density-dependent (open squares), and point-coupling (triangles)
RMF models. The constraints from the isospin diffusion data
(shaded band), the isoscaling data (stars), and the isotopic
dependence of the GMR in even-A Sn isotopes (dashed rectangle with
$L$ constrained by the isospin diffusion data) are also
included.}} \label{LKasyCorr}
\end{figure}

We also see from Table \ref{Bulk} that only five parameter sets,
i.e., TM1, NL$\rho $, NL$\rho \delta $, PKDD and FKVW, in the $23$
parameter sets have nuclear symmetry energies that are consistent
with the extracted values for both $L$ and $ K_{\mathrm{asy}}$
($-500\pm 50$ MeV or $-550\pm 100$ MeV). This can be seen more
clearly in Fig. \ref{LKasyCorr} where the correlation between $L$
and $K_{\mathrm{asy}}$ is displayed for the $23$ parameter sets
together with the constraints from the isospin diffusion data, the
isoscaling data, and the isotopic dependence of the GMR in even-A
Sn isotopes. Fig. \ref{LKasyCorr} further shows that there exists
an approximately linear correlation between $L$ and $
K_{\mathrm{asy}}$, i.e., a larger $L$ leads to a larger magnitude
for $K_{ \mathrm{asy}}$. A similar approximately linear
correlation between $L$ and $K_{ \mathrm{asy}}$ has also been
observed in Ref. \cite{Che05a} for the phenomenological MDI
interactions, and this correlation can be understood from Eq.
(\ref{Kasy2}) which shows that the value of $K_{\mathrm{asy}}$ is
more sensitive to the value of $L$ than to that of
$K_{\mathrm{sym}}$.

The above comparisons thus indicate that the extracted values of
$L=88\pm 25$ MeV and $K_{\mathrm{asy}}=-500\pm 50$ MeV from the
isospin diffusion data, $L\approx 65$ MeV and
$K_{\mathrm{asy}}\approx -453$ MeV from the isoscaling data, and
$K_{\mathrm{asy}}=-550\pm 100$ MeV from the isotopic dependence of
the GMR in even-A Sn isotopes indeed put a very stringent
constraint on the values of the parameters in different RMF
models. The fact that most of the $23$ parameter sets considered
in the present work give symmetry energies that are inconsistent
with the constraints of $L=88\pm 25$ MeV and
$K_{\mathrm{asy}}=-500\pm 50$ MeV or $-550\pm 100$ MeV is probably
related to the rather limited flexibility in the parametrization
of the isovector channel in all RMF models. They are also probably
connected to the fact that most of the parameter sets are obtained
from fitting properties of finite nuclei, which are mostly near
the $\beta $-stability line and thus are not well constrained by
the isospin-dependent properties of nuclear EOS. Also, we are
interested here in the density-dependent behavior of the symmetry
energy around saturation density, as both $L$ and $K_{\mathrm{asy}
} $ are defined at saturation density, while the behavior of the
nuclear EOS at sub-subsaturation density may be more relevant when
the parameter sets are obtained from fitting the properties of
finite nuclei.

\subsection{Nuclear symmetry potential}

\begin{figure}[th]
\includegraphics[scale=1.2]{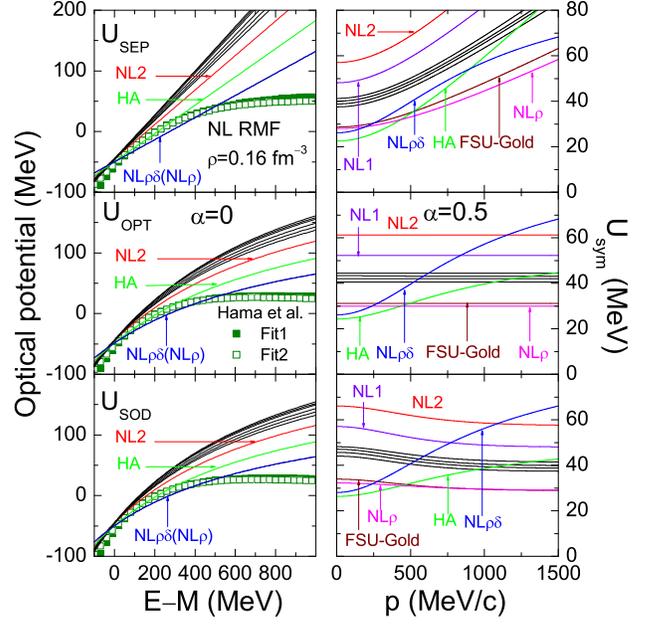}
\caption{{\protect\small (Color online) Energy dependence of the
three different nucleon optical potentials, i.e.,\ }$U_{\text{SEP}}$
{\protect\small \ (Eq. (\protect\ref{Usep})), }$U_{\text{OPT}}$
{\protect\small \ (Eq. (\protect\ref{Uopt})) and }$U_{\text{SOD}}$
{\protect\small \ (Eq. (\protect\ref{Usd})) (left panels) as well as
their corresponding symmetry potentials
}$U_{\text{sym}}^{\text{SEP}}$ {\protect\small , }$U_{
\text{sym}}^{\text{OPT}}${\protect\small , and
}$U_{\text{sym}}^{\text{SOD}}${\protect\small as functions of
momentum (right panels), at a fixed baryon density }$\protect\rho
_{B}=0.16$ {\protect\small \ fm}$^{-3}${\protect\small \ for the
parameter sets NL1, NL2, NL3, NL-SH, TM1, PK1, FSU-Gold, HA,
NL}$\protect\rho ${\protect\small , and NL}$\protect\rho
\protect\delta ${\protect\small \ in the nonlinear RMF model. For
comparison, the energy dependence of the real part of the optical
potential in symmetric nuclear matter at saturation density
extracted from two different fits of the proton-nucleus scattering
data in the Dirac phenomenology are also included (left panels).}}
\label{UpLabNL}
\end{figure}

\begin{figure}[th]
\includegraphics[scale=1.2]{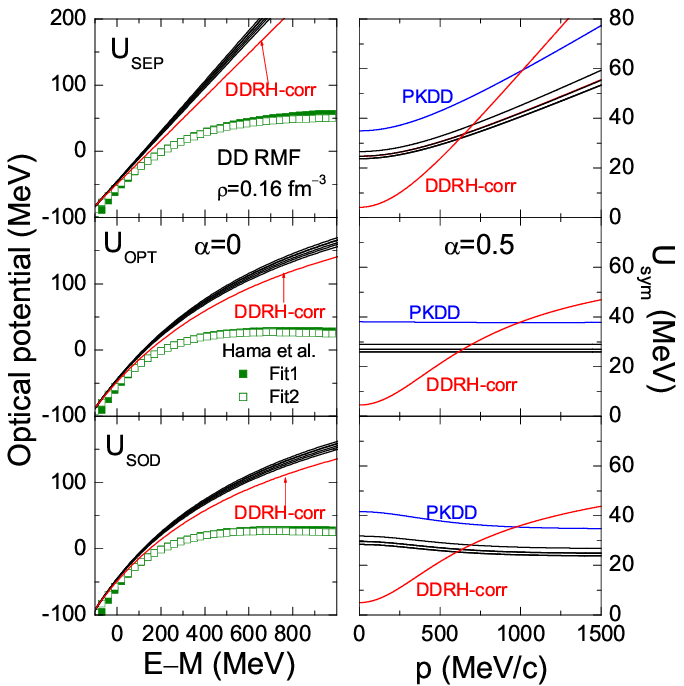}
\caption{{\protect\small (Color online) Same as Fig.
\protect\ref{UpLabNL} for TW99, DD-ME1, DD-ME2, PKDD, DD, DD-F, and
DDRH-corr in the density-dependent RMF models.}} \label{UpLabDD}
\end{figure}

\begin{figure}[th]
\includegraphics[scale=1.2]{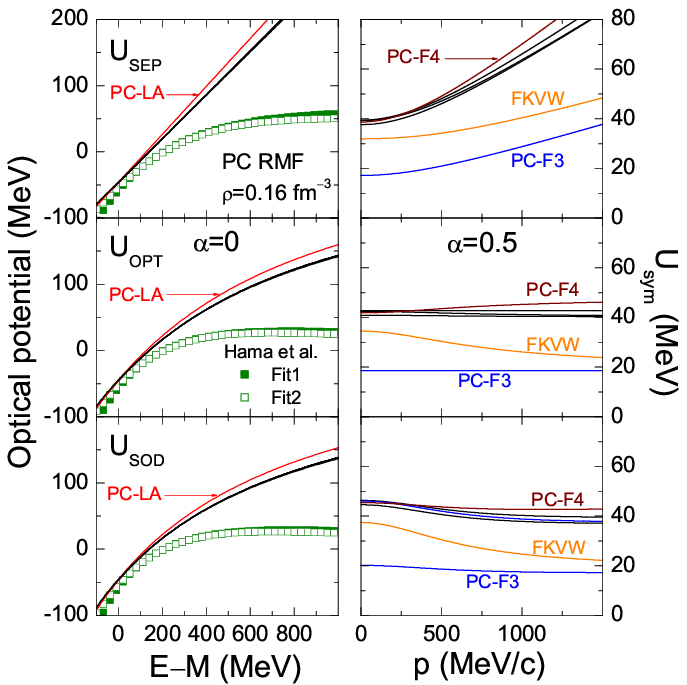}
\caption{{\protect\small (Color online) Same as Fig.
\protect\ref{UpLabNL} for PC-F1, PC-F2, PC-F3, PC-F4, PC-LA, and
FKVW in the point-coupling RMF models.}} \label{UpLabPC}
\end{figure}

Using the parameter sets NL1, NL2, NL3, NL-SH, TM1, PK1, FSU-Gold,
HA, NL$\rho $, and NL$\rho \delta $ \ in the nonlinear RMF model,
we have evaluated the energy dependence of the three different
nucleon optical potentials, i.e., the \textquotedblleft
Schr\"{o}dinger-equivalent potential\textquotedblright\
$U_{\mathrm{SEP}}$ (Eq. (\ref{Usep})), the optical potential from
the difference between the total energy of a nucleon in nuclear
medium and its energy at the same momentum in free space
$U_{\mathrm{OPT}}$ (Eq. (\ref{Uopt})), and the optical potential
based on the second-order Dirac equation $U_{\mathrm{SOD}}$ (Eq.
(\ref{Usd} )), at a fixed baryon density $\rho _{B}=0.16$
fm$^{-3}$ (roughly corresponding to the saturation densities
obtained from various RMF models). For their corresponding
symmetry potentials $U_{\mathrm{sym}}^{ \mathrm{SEP}}$,
$U_{\mathrm{sym}}^{\mathrm{OPT}}$, and $U_{\mathrm{sym}}^{
\mathrm{SOD}}$, we have evaluated instead their dependence on the
nucleon momentum in asymmetric nuclear matter at baryon density
$\rho _{B}=0.16~{\rm fm}^{-3}$ and with isospin asymmetry
$\alpha=0.5$. We note that in contrast to the energy dependence of
the nuclear symmetry potential, the momentum dependence of the
nuclear symmetry potential is almost independent of the isospin
asymmetry of nuclear matter. These results are shown in Fig.
\ref{UpLabNL}. Corresponding results for the parameter sets TW99,
DD-ME1, DD-ME2, PKDD, DD, DD-F, and DDRH-corr in the
density-dependent RMF model and for PC-F1, PC-F2, PC-F3, PC-F4,
PC-LA, and FKVW in the point-coupling RMF model are shown in Figs.
\ref{UpLabDD} and \ref{UpLabPC}, respectively. For comparison, we
also include in these figures results for the energy dependence of
the real part of the different optical potentials in symmetric
nuclear matter at saturation density that are extracted from the
proton-nucleus scattering data based on the Dirac phenomenology
\cite{Ham90,Coo93}.

It is seen that different optical potentials in symmetric nuclear
matter at $\rho _{B}=0.16$ fm$^{-3}$ exhibit similar energy
dependence at low energies but have different behaviors at high
energies. In particular, at high energies, $U_{\mathrm{SEP}}$
continues to increase linearly with energy while
$U_{\mathrm{OPT}}$ and $U_{\mathrm{SOD}}$ seem to saturate at high
energies and thus display a more satisfactory high-energy limit,
similar to what is observed in the nuclear optical potential that
is extracted from the experimental data based on the Dirac
phenomenology. The critical energy at which the optical potential
changes from negative to positive values is between about $130$
MeV and $270$ MeV, depending on the parameter sets used. These
features are easy to understand from the fact that the scalar and
vector potentials are momentum/energy-independent in the RMF
models considered here. Analysis of experimental data from the
proton-nucleus scattering in the Dirac phenomenology also
indicates that the extracted different nucleon optical potentials
in symmetric nuclear matter at normal nuclear density change from
negative to positive values at a nucleon energy of about $208$
MeV. Furthermore, it is seen that the different optical potentials
from all $23$ parameter sets are consistent with the experimental
data at lower energies, i.e., $E_{\mathrm{kin}}<100-200~{\rm
MeV}$, but are generally too repulsive at higher energies,
especially for the \textquotedblleft Schr\"{o}dinger-equivalent
potential\textquotedblright\ $U_{\mathrm{SEP}}$. These features
imply that the RMF models with parameters fitted to the properties
of finite nuclei can only give reasonable description of the low
energy behavior of the isoscalar optical potentials. On the other
hand, it should be mentioned that for optical potentials at high
energies, contributions from dispersive processes such as
dynamical polarization by inelastic excitations, inelastic isobar
resonance excitation above the pion threshold, and particle
production become important \cite{Are02,Fuc06b}. Including such
continuum excitations is expected to improve significantly the
high energy behavior of the optical potential \cite{Are02}. Such
studies are, however, beyond the RMF model based on the Hartree
level as considered here.

For the momentum dependence of the symmetry potential, all $23$
parameter sets display similar behaviors in
$U_{\mathrm{sym}}^{\mathrm{SEP}}$, i.e., increasing with momentum,
albeit at different rates. This can be qualitatively understood as
follows. Expressing Eq. (\ref{Usep}) as
\begin{equation}
U_{\mathrm{SEP},\tau }=\frac{1}{2M_{\tau }}[E_{\tau }^{2}-(M_{\tau
}^{2}+ \vec{p}^{2})],
\end{equation}
and neglecting the difference in neutron and proton masses, we can
rewrite Eq. (\ref{UsymSEP}) as
\begin{eqnarray}
U_{\mathrm{sym}}^{\mathrm{SEP}} &=&\frac{E_{n}^{2}-E_{p}^{2}}{4M_{\tau
}\alpha }  \notag \\
&=&\frac{1}{4M_{\tau }\alpha }[(\Sigma _{n}^{0})^{2}+2\Sigma _{n}^{0}\sqrt{%
\vec{p}^{2}+(M_{n}+\Sigma _{n}^{S})^{2}}  \notag \\
&&+(M_{n}+\Sigma _{n}^{S})^{2}-(\Sigma _{p}^{0})^{2}  \notag \\
&&-2\Sigma _{p}^{0}\sqrt{\vec{p}^{2}+(M_{p}+\Sigma _{p}^{S})^{2}}
-(M_{p}+\Sigma _{p}^{S})^{2}]  \notag \\
&=&\frac{1}{4M_{\tau }\alpha }[(\Sigma _{n}^{0})^{2}-(\Sigma
_{p}^{0})^{2}+(M_{\mathrm{Dirac},n}^{\ast})^{2}  \notag \\
&&-(M_{\mathrm{Dirac},p}^{\ast})^{2}+2\Sigma_{n}^{0}\sqrt{\vec{p}^{2} +(M_{%
\mathrm{Dirac},n}^{\ast})^{2}}  \notag \\
&&-2\Sigma_{p}^{0}\sqrt{\vec{p}^{2}+(M_{\mathrm{Dirac},p}^{\ast })^{2}}].
\end{eqnarray}
In the simple case of the nonlinear RMF model without the
isovector-scalar $\delta $ meson, the neutron Dirac mass is the
same as that of proton. In this case, $U_{\mathrm{sym}}^{
\mathrm{SEP}}$ is reduced to
\begin{eqnarray}
U_{\mathrm{sym}}^{\mathrm{SEP}} &=&\frac{1}{4M_{\tau }\alpha
}[(\Sigma
_{n}^{0})^{2}-(\Sigma _{p}^{0})^{2}  \notag \\
&&+2(\Sigma _{n}^{0}-\Sigma _{p}^{0})\sqrt{\vec{p}^{2}+(M_{\mathrm{Dirac}%
}^{\ast })^{2}}].
\end{eqnarray}%
Since it can be shown from Eqs. (\ref{OmgNL}), ( \ref{RhoNL}), and
(\ref{Sig0NL}) that
\begin{equation}
\Sigma _{n}^{0}-\Sigma _{p}^{0}=2\left( \frac{g_{\rho }}{m_{\rho }}\right)
^{2}(\rho _{n}-\rho _{p}),
\end{equation}
we thus have $\Sigma _{n}^{0}>\Sigma _{p}^{0}$ and an increase of
$U_{\mathrm{\ sym}}^{\mathrm{SEP}}$ with the momentum of a nucleon
in neutron-rich nuclear matter. The same argument applies to
density-dependent RMF models and point-coupling models if the
coupling constant $\alpha _{\mathrm{TV}}$ or $G_{TV}$ in the
point-coupling models is positive (at saturation density) so that
the potential energy part of the symmetry energy at saturation
density is also positive.

For $U_{\mathrm{sym}}^{\mathrm{OPT}}$, whether it increases or
deceases with nucleon momentum depends on the isospin splitting of
the nucleon scalar self energy (scalar potential) or Dirac mass in
neutron-rich nuclear matter. This can be seen from
Eq.(\ref{UsymOPT}) if it is re-expressed as
\begin{eqnarray}
U_{\mathrm{sym}}^{\mathrm{OPT}} &=&\frac{E_{n}-E_{p}}{2\alpha }  \notag \\
&=&\frac{1}{2\alpha }(\Sigma _{n}^{0}-\Sigma _{p}^{0}+\sqrt{\vec{p}
^{2}+(M_{n}+\Sigma _{n}^{S})^{2}}  \notag \\
&&-\sqrt{\vec{p}^{2}+(M_{p}+\Sigma _{p}^{S})^{2}})  \notag \\
&=&\frac{1}{2\alpha }[\Sigma _{n}^{0}-\Sigma _{p}^{0}+\sqrt{\vec{p}^{2}+(M_{
\mathrm{Dirac},n}^{\ast })^{2}}  \notag \\
&&-\sqrt{\vec{p}^{2}+(M_{\mathrm{Dirac},p}^{\ast })^{2}}].
\end{eqnarray}
We note that $U_{\mathrm{sym}}^{\mathrm{OPT}}$ increases with
momentum for the parameter sets HA, NL$\rho \delta $, DDRH-corr, and
PC-F4 while the opposite behavior is observed for the parameter sets
PC-F2, PC-LA, and FKVW.

For the momentum dependence of $U_{\mathrm{sym}}^{\mathrm{SOD}}$, it
is similar to that of $U_{\mathrm{sym}}^{\mathrm{OPT}}$ if we
rewrite Eq. (\ref{UsymSOD}) as
\begin{eqnarray}
U_{\mathrm{sym}}^{\mathrm{SOD}} &=&\frac{E_{n}-E_{p}-(M_{\tau }^{2}+\vec{p}%
^{2})(\frac{1}{E_{n}}-\frac{1}{E_{n}})}{4\alpha }  \notag \\
&=&U_{\mathrm{sym}}^{\mathrm{OPT}}/2-\frac{(M_{\tau }^{2}+\vec{p}^{2})(\frac{
1}{E_{n}}-\frac{1}{E_{n}})}{4\alpha }.
\end{eqnarray}
In this case, $U_{\mathrm{sym}}^{\mathrm{SOD}}$ increases with
nucleon momentum for the parameter sets HA, NL$\rho \delta $, and
DDRH-corr while it decreases for other parameter sets considered
here.

In Ref.\cite{Jam89}, it has been argued that it is the
\textquotedblleft Schr\"{o}dinger-equivalent potential
\textquotedblright\ $U_{\mathrm{SEP}}$ (Eq. (\ref{Usep})) and thus
its corresponding symmetry potential $U_{\mathrm{sym}
}^{\mathrm{SEP}}$ that should be compared with the results from
non-relativistic models. As discussed before, the experimental
data indicate that the nuclear symmetry potential at nuclear
matter saturation density, i.e., the Lane potential
$U_{\mathrm{Lane}}$, clearly decreases at low energies (beam
energy $E_{\mathrm{kin}}$ up to about $100$ MeV and corresponding
momentum values ranging from about $300$ MeV/c to $470$ MeV/c),
which is obviously contradictory to the results for
$U_{\mathrm{sym}}^{\mathrm{SEP}}$ from all of the $23$ parameter
sets considered here. On the other hand, $U_{\mathrm{sym}
}^{\mathrm{OPT}}$ and $U_{\mathrm{sym}}^{\mathrm{SOD}}$ for some
parameter sets can decrease with nucleon momentum, which is
qualitatively consistent with experimental results.

For nucleons with momenta less than about $250-300$ MeV/c or
$E_{\mathrm{kin}}<0$, although observed increase of
$U_{\mathrm{sym} }^{\mathrm{SEP}}$ with momentum for all $23$
parameter sets, and $U_{\mathrm{sym} }^{\mathrm{OPT}}$ as well as
$U_{\mathrm{sym}}^{\mathrm{SOD}}$ with some parameter sets, seems
to be consistent with the results from the microscopic DBHF
\cite{Fuc04}, the extended BHF with 3-body forces \cite{Zuo05},
and chiral perturbation theory calculations \cite{Fri05}, i.e.,
the symmetry potential stays as a constant or slightly increases
with momentum before decreasing at high momenta, it fails to
describe the high momentum/energy behaviors of the nuclear
symmetry potential extracted from nucleon-nucleus scattering
experiments and (p,n) charge exchange reactions at beam energies
up to about $100$ MeV.

We note that in studies based on the relativistic impulse
approximation with empirical $NN$ scattering amplitude and the
nuclear scalar and vector densities from the RMF model, the
Schr\"{o}dinger-equivalent nuclear symmetry potential at fixed
baryon density is found to decrease with increasing nucleon energy
in the range of $100\le E_{\mathrm{kin}}\le 400$ MeV \cite{LiZH06b}
and becomes essentially constant once the nucleon kinetic energy is
greater than about $500$ MeV \cite{Che05c}.

\subsection{Nucleon effective mass}

\begin{table}[tbp]
\caption{{\protect\small Values of different nucleon effective
masses, i.e., }$M_{\mathrm{Dirac}}^{\ast }/M${\protect\small ,
}$M_{\mathrm{Landau}}^{\ast }/M${\protect\small ,
}$M_{\mathrm{Lorentz}}^{\ast }/M${\protect\small , }$
M_{\mathrm{OPT}}^{\ast }/M${\protect\small , and
}$M_{\mathrm{SOD}}^{\ast }/M ${\protect\small \ in symmetric nuclear
matter at saturation density using the }$23${\protect\small \
parameter sets in the nonlinear, density-dependent, and
point-coupling RMF models. The last column gives the references for
corresponding parameter sets.}}
\label{EffMass}%
\begin{tabular}{ccccccc}
\hline\hline
Model & $\frac{M_{Dirac}^{\ast }}{M}$ & $\frac{M_{Landau}^{\ast }}{M}$ & $%
\frac{M_{Lorentz}^{\ast }}{M}$ & $\frac{M_{OPT}^{\ast }}{M}$ & $\frac{
M_{SOD}^{\ast }}{M}$ & Ref. \\ \hline
NL1 & $0.57$ & $0.64$ & $0.65$ & $0.61$ & $0.59$ & \cite{Lee86} \\
NL2 & $0.67$ & $0.72$ & $0.74$ & $0.70$ & $0.68$ & \cite{Lee86} \\
NL3 & $0.60$ & $0.66$ & $0.67$ & $0.63$ & $0.61$ & \cite{Lal97} \\
NL-SH & $0.60$ & $0.66$ & $0.67$ & $0.63$ & $0.61$ & \cite{Sha93} \\
TM1 & $0.63$ & $0.69$ & $0.71$ & $0.67$ & $0.65$ & \cite{Sug94} \\
PK1 & $0.61$ & $0.66$ & $0.68$ & $0.64$ & $0.62$ & \cite{Lon04} \\
FSUGold & $0.61$ & $0.67$ & $0.69$ & $0.65$ & $0.62$ & \cite{Tod05} \\
HA & $0.68$ & $0.74$ & $0.75$ & $0.71$ & $0.69$ & \cite{Bun03} \\
NL$\rho $ & $0.75$ & $0.80$ & $0.82$ & $0.77$ & $0.76$ & \cite{Liu02} \\
NL$\rho \delta $ & $0.75$ & $0.80$ & $0.82$ & $0.77$ & $0.76$ & \cite{Liu02}
\\
&  &  &  &  &  &  \\
TW99 & $0.55$ & $0.62$ & $0.64$ & $0.60$ & $0.57$ & \cite{Typ99} \\
DD-ME1 & $0.58$ & $0.64$ & $0.66$ & $0.62$ & $0.59$ & \cite{Nik02} \\
DD-ME2 & $0.57$ & $0.63$ & $0.65$ & $0.61$ & $0.59$ & \cite{Lal05} \\
PKDD & $0.57$ & $0.63$ & $0.65$ & $0.61$ & $0.59$ & \cite{Lon04} \\
DD & $0.56$ & $0.63$ & $0.64$ & $0.61$ & $0.58$ & \cite{Typ05} \\
DD-F & $0.56$ & $0.62$ & $0.64$ & $0.60$ & $0.57$ & \cite{Kla06} \\
DDRH-corr & $0.55$ & $0.63$ & $0.64$ & $0.60$ & $0.58$ & \cite{Hof01} \\
&  &  &  &  &  &  \\
PC-F1 & $0.61$ & $0.67$ & $0.69$ & $0.64$ & $0.62$ & \cite{Bur02} \\
PC-F2 & $0.61$ & $0.67$ & $0.69$ & $0.64$ & $0.62$ & \cite{Bur02} \\
PC-F3 & $0.61$ & $0.67$ & $0.69$ & $0.64$ & $0.62$ & \cite{Bur02} \\
PC-F4 & $0.61$ & $0.67$ & $0.69$ & $0.64$ & $0.62$ & \cite{Bur02} \\
PC-LA & $0.58$ & $0.64$ & $0.65$ & $0.61$ & $0.59$ & \cite{Bur02} \\
FKVW & $0.62$ & $0.68$ & $0.70$ & $0.65$ & $0.63$ & \cite{Fin06} \\
\hline\hline
\end{tabular}%
\end{table}

For the different nucleon effective masses in symmetric nuclear
matter at saturation density, we show in Table \ref{EffMass} the
results from the $23$ parameter sets in the nonlinear,
density-dependent, and point-coupling RMF models. It is seen that
the values of $M_{\mathrm{Dirac} }^{\ast }/M$,
$M_{\mathrm{Landau}}^{\ast }/M$, $M_{\mathrm{Lorentz}}^{\ast }/M$,
$M_{\mathrm{OPT}}^{\ast }/M$, and $M_{\mathrm{SOD}}^{\ast }/M$ are
in the range of $0.55\sim 0.75$, $0.62\sim 0.80$, $0.64\sim 0.80$,
$0.60\sim 0.77$, and $0.57\sim 0.76$, respectively. The parameter
sets NL2, HA, NL$\rho $ and NL$\rho \delta $ seem to give too
large values, i.e., $0.67$, $0.68$, $0.75$, and $0.75$,
respectively, for the $M_{\mathrm{Dirac}}^{\ast }/M$ as values in
the range of $0.55\sim 0.60$ are needed to describe reasonably the
spin-orbit splitting in finite nuclei using the RMF models. On the
other hand, the larger Dirac masses leads to larger Landau masses
$M_{\mathrm{Landau}}^{\ast }/M$ of $0.72$, $0.74$, $0.80$, and $
0.80$, respectively, for the parameter sets NL2, HA, NL$\rho $ and
NL$\rho \delta $, which are consistent with the empirical
constraint of $M_{\mathrm{\ Landau}}^{\ast }/M$ = $0.8\pm 0.1$
\cite{Cha97,Cha98,Rei99,Mar07}.

\begin{figure}[th]
\includegraphics[scale=0.8]{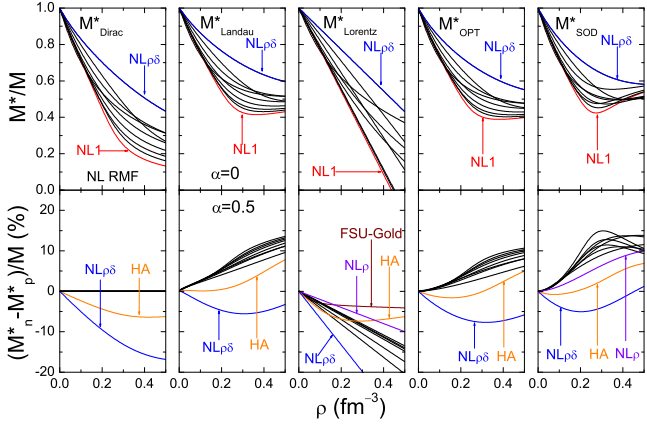}
\caption{{\protect\small (Color online) Density dependence of
different nucleon effective masses, i.e., }$M_{\mathrm{Dirac}}^{\ast
}/M$ {\protect\small , }$M_{\mathrm{Landau}}^{\ast
}/M${\protect\small , }$M_{ \mathrm{Lorentz}}^{\ast
}/M${\protect\small , }$M_{\mathrm{OPT}}^{\ast }/M$ {\protect\small
, and }$M_{\mathrm{SOD}}^{\ast }/M${\protect\small \ in symmetric
nuclear matter as well as their corresponding isospin splittings in
neutron-rich nuclear matter with isospin asymmetry }$\protect\alpha
=0.5$ {\protect\small \ for the parameter sets NL1, NL2, NL3, NL-SH,
TM1, PK1, FSU-Gold, HA, NL}$\protect\rho ${\protect\small , and
NL}$\protect\rho \protect\delta ${\protect\small \ in the nonlinear
RMF model.}} \label{MstarDenNL}
\end{figure}

\begin{figure}[th]
\includegraphics[scale=0.8]{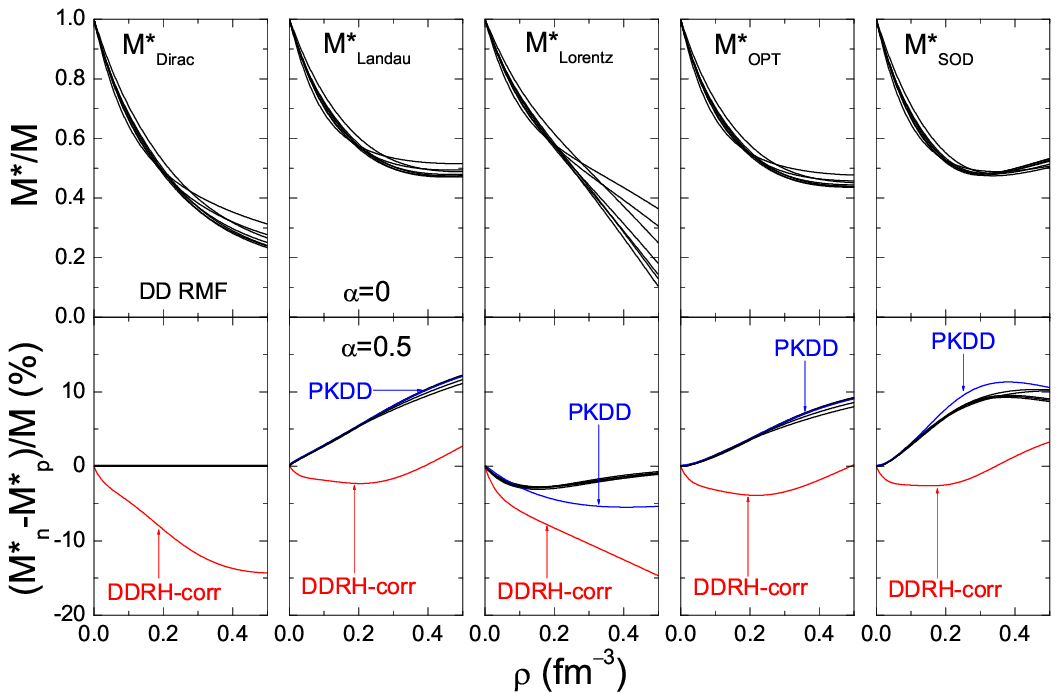}
\caption{{\protect\small (Color online) Same as Fig.
\protect\ref{MstarDenNL} but for TW99, DD-ME1, DD-ME2, PKDD, DD,
DD-F, and DDRH-corr in the density-dependent RMF model.}}
\label{MstarDenDD}
\end{figure}

\begin{figure}[th]
\includegraphics[scale=0.8]{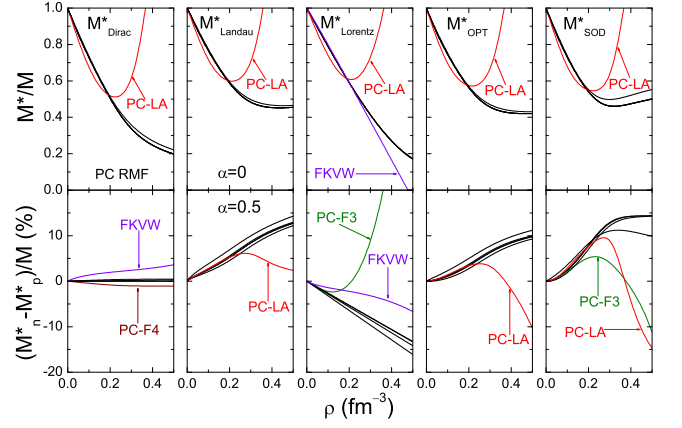}
\caption{{\protect\small (Color online) Same as Fig.
\protect\ref{MstarDenNL} but for PC-F1, PC-F2, PC-F3, PC-F4, PC-LA,
and FKVW in the point-coupling RMF model.}} \label{MstarDenPC}
\end{figure}

The density dependence of the different nucleon effective masses
in symmetric nuclear matter and corresponding isospin splitting
$(M_{n}^{\ast }-M_{p}^{\ast })/M$ in asymmetric nuclear matter
with isospin asymmetry $\alpha =0.5$ are shown in Fig.
\ref{MstarDenNL} for the parameter sets NL1, NL2, NL3, NL-SH, TM1,
PK1, FSU-Gold, HA, NL$\rho $, and NL$\rho \delta $\ in the
nonlinear RMF model. Figs. \ref{MstarDenDD} and \ref{MstarDenPC}
display the same results as in Fig. \ref{MstarDenNL} but for the
parameter sets TW99, DD-ME1, DD-ME2, PKDD, DD, DD-F, and DDRH-corr
in the density-dependent RMF models and for PC-F1, PC-F2, PC-F3,
PC-F4, PC-LA, and FKVW in the point-coupling RMF model,
respectively. It is seen that different parameter sets in the
nonlinear RMF model give significantly different density
dependence for the nucleon effective masses while the different
parameter sets in the density-dependent and point-coupling RMF
models predict roughly the same density dependence for the nucleon
effective masses except that the parameter set PC-LA gives very
large values for the nucleon effective masses at high densities.
This unusual behavior for PC-LA was also observed in Ref.
\cite{Bur02}, and it is due to the fact that the coupling constant
$\gamma _{\mathrm{S}}$ for the higher-order interaction term in
PC-LA is positive \cite{Nik92} and dominates at high density,
leading thus to the very large nucleon effective mass.

For the Landau mass at a fixed baryon density, its value
$M_{\mathrm{Landau}}^{\ast }/M$ is generally larger than
$M_{\mathrm{Dirac}}^{\ast }/M$. This can be seen from Eq.
(\ref{MLandau}) if it is rewritten as
\begin{eqnarray}
M_{\mathrm{Landau},\tau }^{\ast } &=&(E_{\tau }-\Sigma _{\tau }^{0})=\sqrt{%
p_{F\text{,}\tau }^{2}+(M_{\tau }+\Sigma _{\tau }^{S})^{2}}  \notag \\
&=&\sqrt{p_{F,\tau }^{2}+M_{\mathrm{Dirac},\tau }^{\ast 2}}  \label{MLanDir}
\end{eqnarray}
which shows that $M_{\mathrm{Landau},\tau }^{\ast }\geq
M_{\mathrm{Dirac},\tau }^{\ast }$ if nucleon self-energies are
independent of momentum/energy.

For the Lorentz mass $M_{\mathrm{Lorentz}}^{\ast }$, $M_{\mathrm{\
Lorentz}}^{\ast }/M$ depends almost linearly on density and thus has
a stronger density dependence than the Dirac and Landau masses. We
note from Eqs. (\ref{dispersion}) and (\ref{MLanDir}) that Eq.
(\ref{MLorentz}) can be reduced to
\begin{equation}
M_{\mathrm{Lorentz},\tau }^{\ast }=M_{\tau }-\Sigma _{\tau }^{0},
\label{MLorentz2}
\end{equation}
if nucleon self-energies are independent of momentum/energy.
Therefore, the density dependence of $M_{\mathrm{Lorentz}}^{\ast
}$ is determined uniquely by the density dependence of nucleon
vector self-energy. In the nonlinear RMF model, most of the
parameter sets, except for TM1, PK1 and FSU-Gold which include the
self-coupling of the $\omega $ meson field, give a linear density
dependence for $\Sigma _{\tau }^{0}$, leading thus to the observed
linear density dependence of $M_{\mathrm{Lorentz}}^{\ast }$. As to
the nonlinear density dependence of $M_{\mathrm{Lorentz}}^{\ast }$
in the density-dependent RMF model and point-coupling models, it
is due to the nonlinear density dependence of the coupling
constant or the inclusion of higher-order couplings.

For $M_{\mathrm{OPT}}^{\ast }/M$ and $M_{\mathrm{\ SOD}}^{\ast
}/M$, they are seen to have roughly same magnitude and also same
density dependence as $M_{\mathrm{Landau}}^{\ast }/M$. This
feature can be understood from the fact that with the dispersion
relation of Eq. (\ref{dispersion}), Eq. (\ref{Mopt}) and Eq.
(\ref{Msod}) can be re-expressed as
\begin{equation}
M_{\mathrm{OPT},\tau }^{\ast }=\frac{M_{\tau }}{\sqrt{p_{F,\tau
}^{2}+M_{\tau }^{2}}}M_{\mathrm{Landau},\tau }^{\ast }  \label{Mopt2}
\end{equation}
and
\begin{equation}
M_{\mathrm{SOD},\tau }^{\ast }=M_{\tau
}\left[\frac{M_{\mathrm{Landau},\tau }^{\ast }}{E_{\tau
}}+\frac{E_{\tau }^{2}-(p_{F,\tau }^{2}+M_{\tau }^{2})}{ 2E_{\tau
}^{2}}\right], \label{Msod2}
\end{equation}
respectively. Since $p_{F\text{,}\tau }^{2}\ll M_{\tau }^{2}$ (For
example, $p_{F}\approx 385$ MeV/c at $\rho _{B}=0.5$ fm$^{-3}$),
we have $M_{\tau }/\sqrt{p_{F,\tau }^{2}+M_{\tau }^{2}}\approx 1$
(with an error of a few percent) and thus $M_{\mathrm{OPT},\tau
}^{\ast }\approx M_{\mathrm{Landau},\tau }^{\ast }$. Furthermore,
the second term in Eq. (\ref{Msod2}) can be neglected compared
with the first term as $M_{\tau }/E_{\tau }\sim 1$ (it is a good
approximation at low densities and with an error of about $20\%$
at high densities, e.g., $\rho _{B}=0.5$ fm$^{-3}$). As a result,
we have $M_{\mathrm{SOD},\tau }^{\ast }\sim
M_{\mathrm{Landau},\tau }^{\ast }$.

From the Dirac equation, one sees that condensed scalar fields
(scalar self-energies) lead to a shift of nucleon mass such that
the nuclear matter is described as a system of pseudo-nucleons
with masses $M^{\ast }$ (Dirac mass) moving in classical vector
fields with $\delta $ meson field or isovector-scalar potential
further generating the splitting of the proton and neutron Dirac
masses in asymmetric nuclear matter. For the isospin splitting of
$M_{\mathrm{Dirac}}^{\ast }$ in neutron-rich nuclear matter, it is
interesting to see that the parameter sets HA, NL$\rho \delta $,
DDRH-corr, and PC-F4 give $M_{\mathrm{Dirac},p}^{\ast
}>M_{\mathrm{Dirac},n}^{\ast }$ while PC-F2, PC-LA, and FKVW
exhibit the opposite behavior of $M_{\mathrm{\ Dirac},p}^{\ast
}<M_{\mathrm{Dirac},n}^{\ast }$. This feature implies that the
isospin-dependent scalar potential can be negative or positive
depending on the parameter sets used. In the nonlinear RMF model,
we obtain from Eqs. (\ref{DelNL}) and (\ref{MDiracNL})
\begin{equation}
M_{\mathrm{Dirac},n}^{\ast }-M_{\mathrm{Dirac},p}^{\ast }=-2\left( \frac{
g_{\delta }}{m_{\delta }}\right) ^{2}(\rho _{S,n}-\rho _{S,p}),
\end{equation}
which indicates that we always have $M_{\mathrm{Dirac},p}^{\ast
}>M_{\mathrm{\ Dirac},n}^{\ast }$ in the neutron-rich nuclear matter
where $\rho _{S,n}>\rho _{S,p}$. This argument is also applicable to
the density-dependent RMF model by replacing $g_{\delta }$ with the
density dependent $\Gamma _{\delta }$. For the nonlinear
point-coupling models, we have, on the other hand,
\begin{equation}
M_{\mathrm{Dirac},n}^{\ast }-M_{\mathrm{Dirac},p}^{\ast }=2\alpha _{\mathrm{%
\ TS}}(\rho _{S,n}-\rho _{S,p}).
\end{equation}
A similar equation can be obtained for the density-dependent
point-coupling models with the replacement of $\alpha
_{\mathrm{TS}}$ by the density dependent $G_{TS}$. Therefore, the
isospin splitting of $M_{\mathrm{Dirac}}^{\ast }$ in neutron-rich
nuclear matter depends on the sign of the isovector-scalar coupling
constant $\alpha _{\mathrm{TS}}$ and $G_{TS}$ in the point-coupling
models. Since the value of $\alpha _{ \mathrm{TS}}$ in PC-F2 and
PC-LA as well as the value of $G_{TS}$ in FKVW are positive, these
parameter sets lead to the isospin-splitting
$M_{\mathrm{Dirac},n}^{\ast }>M_{ \mathrm{Dirac},p}^{\ast }$ in
neutron-rich nuclear matter, which is opposite to that in other
parameter sets considered here. The isospin splitting of
$M_{\mathrm{Dirac}}^{\ast }$ is directly related to the isovector
spin-orbit potential that determines the isospin-dependent
spin-orbit splitting in finite nuclei. Unfortunately, there are no
clear experimental indication about the isospin dependence of the
spin-orbit splitting in finite nuclei \cite{Hof01}, so detailed
experimental data on the single-particle energy levels in exotic
nuclei are needed to pin down the isospin splitting of
$M_{\mathrm{Dirac}}^{\ast }$ in asymmetric nuclear matter.

For the isospin splitting of $M_{\mathrm{Landau}}^{\ast }$ in
neutron-rich nuclear matter, most parameter sets give
$M_{\mathrm{Landau},n}^{\ast }>M_{\mathrm{Landau},p}^{\ast }$,
which is consistent with the usual results in non-relativistic
models. The parameter sets NL$\rho \delta $ and DDRH-corr give,
however, the opposite result due to the strong isospin-splitting
of $M_{\mathrm{Dirac}}^{\ast }$ with $M_{\mathrm{Dirac},n}^{\ast
}<M_{\mathrm{\ Dirac},p}^{\ast }$ for NL$\rho \delta $ and
DDRH-corr and the fact that $M_{\mathrm{Landau}}^{\ast }$ is
related to the Fermi momentum and $M_{\mathrm{Dirac}}^{\ast }$
according to Eq. (\ref{MLanDir}). The isospin-splitting
$M_{\mathrm{Landau},n}^{\ast }>M_{ \mathrm{Landau},p}^{\ast }$
implies that neutrons have a larger level density at the Fermi
energy and thus more compressed single-particle levels in finite
nuclei than protons.

For the isospin splitting of $M_{\mathrm{Lorentz}}^{\ast }$ in
neutron-rich nuclear matter, all parameter sets give
$M_{\mathrm{Lorentz},p}^{\ast }>M_{\mathrm{Lorentz},n}^{\ast }$
except that the PC-L3 gives $M_{\mathrm{Lorentz },p}^{\ast
}<M_{\mathrm{Lorentz},n}^{\ast }$ at high densities. From Eq.
(\ref{MLorentz2}), we have
\begin{equation}
M_{\mathrm{Lorentz},n}^{\ast }-M_{\mathrm{Lorentz},p}^{\ast }=-(\Sigma
_{n}^{0}-\Sigma _{p}^{0}),
\end{equation}
which leads to the observed isospin-splitting $M_{\mathrm{Lorentz}
,p}^{\ast }>M_{\mathrm{Lorentz},n}^{\ast }$ as we generally have
$\Sigma _{n}^{0}>\Sigma _{p}^{0}$ as discussed above. For the
parameter set PC-L3, it includes a higher-order isovector-vector
term through the parameter $\gamma _{\mathrm{TV}}$. Since the
latter has a negative value and dominates at high densities
according to Eq. (\ref{Sig0NLPC}), it leads to $\Sigma
_{n}^{0}<\Sigma _{p}^{0}$ and thus $M_{\mathrm{Lorentz},p}^{\ast
}<M_{\mathrm{Lorentz},n}^{\ast }$ at high densities. The isospin
splitting of $M_{\mathrm{OPT}}^{\ast }/M$ and
$M_{\mathrm{SOD}}^{\ast }/M$ in neutron-rich nuclear matter show a
similar behavior as $M_{\mathrm{Landau}}^{\ast }$ as expected from
the discussions below Eqs. (\ref{Mopt2}) and (\ref{Msod2}).

\subsection{Nucleon scalar density}

The nucleon scalar density as defined in Eq. (\ref{RhoS}) is the
source for the nucleon scalar self-energy (scalar potential). In
the RMF model, the isospin-dependent nucleon scalar density is
uniquely related to the nucleon Dirac mass as shown in Eq.
(\ref{RhoSnp}). The latter equation also shows that the scalar
density is less than the baryon density due to the factor
$M_{i}^{\ast}/\sqrt{\vec{k}^{2}+(M_{i}^{\ast })^{2}}$ which causes
a reduction of the contribution of rapidly moving nucleons to the
scalar source term. This mechanism is responsible for nuclear
matter saturation in the mean-field theory and essentially
distinguishes relativistic models from non-relativistic ones. In
practice, the isospin-dependent nucleon scalar density is also an
essential ingredient for evaluating the relativistic optical
potential for neutrons and protons in the relativistic impulse
approximation (See, e.g., Refs. \cite{Che05c,LiZH06b} and
references therein).

\begin{figure}[th]
\includegraphics[scale=1.2]{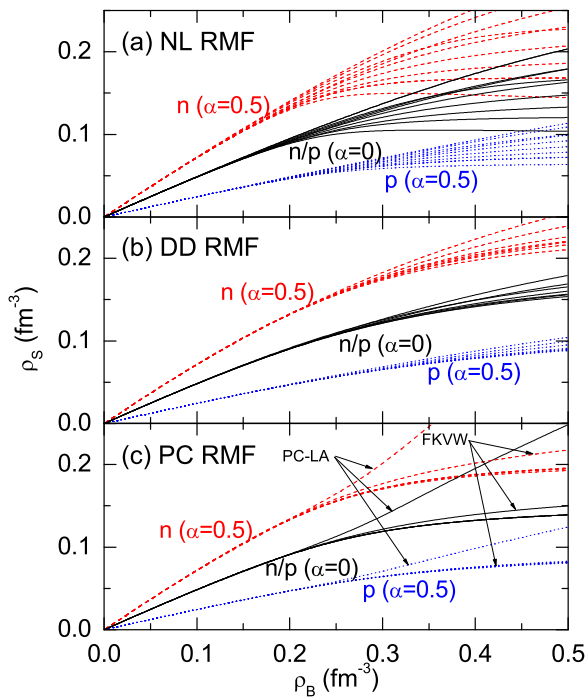}
\caption{{\protect\small (Color online) Neutron and proton scalar densities
as functions of baryon density in nuclear matter with isospin asymmetry }$%
\protect\alpha =0${\protect\small \ and }$0.5${\protect\small \ for the
parameter sets NL1, NL2, NL3, NL-SH, TM1, PK1, FSU-Gold, HA, NL}$\protect%
\rho ${\protect\small , and NL}$\protect\rho \protect\delta ${\protect\small %
\ of the nonlinear RMF model (a); TW99, DD-ME1, DD-ME2, PKDD, DD, DD-F, and
DDRH-corr of the density-dependent RMF model (b); PC-F1, PC-F2, PC-F3,
PC-F4, PC-LA, and FKVW of the point-coupling RMF model (c).}}
\label{RhoSRhoB}
\end{figure}

In Fig. \ref{RhoSRhoB}, we show the neutron and proton scalar
densities as functions of the baryon density $\rho _{B}$ in
nuclear matter with isospin asymmetry $\alpha =0$ and $0.5$ for
the $23$ parameter sets from the nonlinear, density-dependent, and
point-coupling RMF models. It is seen that the neutron scalar
density is larger than that of protons in neutron-rich nuclear
matter at a fixed baryon density. Although results for different
parameter sets are almost the same at lower baryon densities, they
become different when $\rho _{B}\gtrsim 0.25$ fm$^{-3}$, and this
is consistent with the conclusions of Refs. \cite{Che05c,LiZH06b}.
In particular, different parameter sets in the nonlinear RMF model
predict a larger uncertainty for the value of the nucleon scalar
density at high baryon density while all the parameter sets
(except PC-LA) in the density-dependent RMF model and
point-coupling models give roughly same results for the nucleon
scalar density. These features are consistent with the results for
the density dependence of nucleon Dirac mass shown in Figs.
\ref{MstarDenNL}, \ref{MstarDenDD}, and \ref{MstarDenPC}. At low
baryon densities, neutron and proton scalar densities are seen to
increase roughly linearly with baryon density, and this can be
easily understood from Eq. (\ref{RhoSnp}), which is reduced to the
following expression at low densities ($|\vec{k}|\rightarrow 0$
due to $ k_{F}\rightarrow 0$):
\begin{eqnarray}
\rho _{S,i} &\approx &\frac{2}{{(2\pi )}^{3}}\int_{0}^{k_{F}^{i}}d^{3}\!k\,
\frac{M_{i}^{\ast }}{M_{i}^{\ast }}  \notag \\
&=&\frac{2}{{(2\pi )}^{3}}\int_{0}^{k_{F}^{i}}d^{3}\!k\,=\rho _{B,i},\text{ }
i=p,n.
\end{eqnarray}
Therefore, neutron and proton scalar densities generally approach
their respective baryon densities in asymmetric nuclear matter at
low baryon densities.

\section{Summary and conclusions}

\label{summary}

Using different versions of relativistic mean-field models that
are commonly used in current nuclear structure studies, i.e., the
nonlinear model, the model with density-dependent nucleon-meson
coupling, and the point-coupling model, we have investigated
systematically the isospin-dependent bulk and single-particle
properties of isospin-asymmetric nuclear matter. In particular, we
considered $23$ parameter sets commonly and successfully used in
nuclear structure studies, i.e., NL1, NL2, NL3, NL-SH, TM1, PK1,
FSU-Gold, HA, NL$\rho $, NL$\rho \delta $ for the nonlinear RMF
model; TW99, DD-ME1, DD-ME2, PKDD, DD, DD-F, and DDRH-corr for the
density-dependent RMF model; and PC-F1, PC-F2, PC-F3, PC-F4,
PC-LA, and FKVW for the point-coupling RMF model. Most of the
parameter sets are obtained from fitting the binding energies and
charge radii of a large number of nuclei in the periodic table or
the results from the microscopic DBHF approach, which have been
shown to describe successfully a number of the properties of
finite nuclei.

Using these models, we have studied the density dependence of
nuclear symmetry energy and compared the results with the symmetry
energy recently extracted from the analyses of the isospin
diffusion data from heavy-ion collisions based on an isospin- and
momentum-dependent transport model with in-medium NN cross
sections, the isoscaling analyses of isotope ratios in
intermediate energy heavy ion collisions, and measured isotopic
dependence of the giant monopole resonances in even-A Sn isotopes.
These analyses have led to the extraction of $L=88\pm 25$ MeV for
the slope parameter of the nuclear symmetry energy at saturation
density and $K_{ \mathrm{asy}}=-500\pm 50$ MeV or $-550\pm 100$
MeV for the isospin-dependent part of the isobaric
incompressibility of isospin asymmetric nuclear matter, which may
represent the most stringent phenomenological constraints
available so far on the nuclear symmetry energy at sub-saturation
densities. Using these constraints, we have found that, among the
$23$ parameter sets considered in the present work, only six sets,
i.e., TM1, NL$\rho $, NL$\rho \delta $, PKDD, PC-LA, and FKVW,
have nuclear symmetry energies that are consistent with the
extracted $L$ value of $88\pm 25$ MeV while fifteen sets, i.e.,
NL3, NL-SH, TM1, PK1, HA, NL$\rho $, NL$\rho \delta $, TW99, PKDD,
DD-F, PC-F1, PC-F2, PC-F3, PC-F4, and FKVW, have nuclear symmetry
energies that are consistent with the extracted $K_{\mathrm{asy}}$
value of $-500\pm 50$ MeV or $-550\pm 100$ MeV. Furthermore, we
have found surprisingly that only five parameter sets, i.e., TM1,
NL$\rho $, NL$\rho \delta $, PKDD, and FKVW, in the $23$ parameter
sets have nuclear symmetry energies that are consistent with the
extracted values for both $L$ and $K_{\mathrm{asy}}$. We have
noted that most parameter sets in the nonlinear and point-coupling
RMF models predict stiffer symmetry energies while those in the
density-dependent RMF model give softer symmetry energies. These
features are probably related to the rather limited flexibility in
the parametrization of the isovector channel in all RMF models and
also the fact that most of the parameter sets are obtained from
fitting properties of finite nuclei which are mostly near the
$\beta $-stability line and thus have less constraint on the
isospin-dependent properties of asymmetric nuclear matter.
Moreover, we have focused here on the behavior of the symmetry
energy around saturation density while the parameter sets in RMF
models are fitted to the properties of finite nuclei that are more
sensitive to the properties of the nuclear symmetry energy at
sub-saturation densities.

We have also investigated the energy  dependence of three
different nucleon optical potentials, i.e., the \textquotedblleft
Schr\"{o} dinger-equivalent potential\textquotedblright\
$U_{\mathrm{SEP}}$ (Eq. (\ref{Usep})), the optical potential from
the difference between the total energy of a nucleon in nuclear
medium and its energy at the same momentum in free space
$U_{\mathrm{OPT}}$ (Eq. (\ref{Uopt})), and the optical potential
based on the second-order Dirac equation $U_{\mathrm{SOD}}$ (Eq.
(\ref{Usd} )), as well as their corresponding symmetry potentials
$U_{\mathrm{sym}}^{ \mathrm{SEP}}$,
$U_{\mathrm{sym}}^{\mathrm{OPT}}$, and $U_{\mathrm{sym}}^{
\mathrm{SOD}}$ as functions of momentum. The results indicate that
different optical potentials in symmetric nuclear matter exhibit
similar energy dependence at low energies but have different high
energy behaviors. In particular, at high energies,
$U_{\mathrm{SEP}}$ continues to increase linearly with momentum
while $U_{\mathrm{OPT}}$ and $U_{\mathrm{SOD}}$ seem to saturate
and thus display a more satisfactory high-energy limit compared to
the optical potentials extracted from proton-nucleus scatterings
using the Dirac phenomenology. On the other hand, the nuclear
symmetry potential at a fixed baryon density can increase or
decrease with increasing nucleon momentum depending on the
definition for the nucleon optical potential and the interactions
used. For $U_{\mathrm{sym}}^{\mathrm{SEP}}$ at $\rho _{B}=0.16$
fm$^{-3}$, results from all $23$ parameter sets show that it
increases with momentum, which is consistent with the predictions
of microscopic DBHF and chiral perturbation calculations at low
momenta (less than about 300 MeV/c) but is inconsistent with the
experimental result that the nuclear symmetry potential at
saturation density (the Lane potential $U_{\mathrm{Lane}}$)
decreases at low energies (beam energy $E_{\mathrm{kin}}$ above
$0$ MeV and less than about $100$ MeV and corresponding momentum
values are from about $300$ MeV/c to $470$ MeV/c) and RIA
predictions at higher energies. For
$U_{\mathrm{sym}}^{\mathrm{OPT}}$ and
$U_{\mathrm{sym}}^{\mathrm{SOD}}$, they can, however, decrease
with momentum for some parameter sets, which is qualitatively
consistent with the experimental constraint. Again, we emphasize
that, for the three definitions of the optical potential and thus
their corresponding nuclear symmetry potentials, only
$U_{\mathrm{SEP},\tau }$ is well-defined theoretically and is
Schr\"{o}dinger-equivalent while $U_{\mathrm{OPT},\tau }$ and
$U_{\mathrm{SOD},\tau }$ are used here for references as
$U_{\mathrm{OPT},\tau }$ has been extensively used in microscopic
DBHF calculations \cite{LiGQ93} and transport models for heavy-ion
collisions \cite{Dan00} and $U_{\mathrm{SOD},\tau }$ has been used
in analyses of the relativistic optical potential based on the
Dirac phenomenology \cite{Ham90}.

We have further explored different nucleon effective masses, i.e.,
$M_{\mathrm{Dirac}}^{\ast }$, $M_{\mathrm{Landau}}^{\ast }$,
$M_{\mathrm{\ Lorentz}}^{\ast }$, $M_{\mathrm{OPT}}^{\ast }$, and
$M_{\mathrm{SOD}}^{\ast } $ in symmetric nuclear matter as well as
their isospin-splittings in neutron-rich nuclear matter. Most of
the parameter sets are found to give reasonable values for
$M_{\mathrm{Dirac}}^{\ast }$ as required by the spin-orbit
splitting data in finite nuclei but too small values for
$M_{\mathrm{Landau} }^{\ast }$, implying that they would give too
small a level density at the Fermi energy and too large a spread
of the single-particle levels in finite nuclei. For
$M_{\mathrm{Lorentz}}^{\ast }$, it is found to display the
strongest (almost linear) density dependence even at high
densities. Interestingly, including the isovector-scalar channel
leads to the isospin-splitting of $M_{\mathrm{Dirac}}^{\ast }$,
and $M_{\mathrm{\ Dirac},n}^{\ast }>M_{\mathrm{Dirac},p}^{\ast }$
is always obtained in neutron-rich nuclear matter for the
nonlinear and density-dependent RMF models but an opposite result
can be observed in the point-coupling model. For
$M_{\mathrm{Landau}}^{\ast }$, most parameter sets give the
isospin splitting $M_{\mathrm{Landau},n}^{\ast
}>M_{\mathrm{Landau},p}^{\ast }$ in neutron-rich nuclear matter,
which is consistent with usual results in non-relativistic models,
while an opposite isospin-splitting is observed for
$M_{\mathrm{Lorentz}}^{\ast }$. In addition,
$M_{\mathrm{OPT}}^{\ast }$, and $M_{\mathrm{SOD}}^{\ast }$ are
found to display similar behaviors as $M_{ \mathrm{Landau}}^{\ast
}$.

Finally, we have studied the baryon density dependence of the
nucleon scalar density and its isospin-splitting in neutron-rich
nuclear matter. The results indicate that the neutron scalar
density is larger than that of proton in neutron-rich nuclear
matter at a fixed baryon density. At low baryon densities, the
neutron and proton scalar densities generally approach their
respective baryon densities in asymmetric nuclear matter.

In the present work, we have focused on three versions of standard
RMF models, i.e., the nonlinear, density-dependent, and
point-coupling RMF models. We note that there are some recent
works \cite{Liu04,Ava06,Khv07,Jia07a,Jia07b} in which the standard
RMF models are extended to include density-dependent hadron masses
and meson coupling constants via the Brown-Rho (BR) scaling
\cite{Bro91}. In particular, the parameter sets SLC and SLCd
constructed in Ref. \cite{Jia07a,Jia07b} are not only consistent
with current experimental results for symmetric matter at normal
and supra-normal densities and the symmetry energy constrained by
the isospin diffusion data at sub-saturation densities, but also
give a fairly satisfactory description of the ground state
properties of finite nuclei, including binding energies, charge
radii, and neutron skin thickness.

In all standard RMF models, the nucleon self-energies are
independent of momentum/energy. As a result, the Dirac mass and
the Landau mass obtained from these models cannot be
simultaneously consistent with experimental data (see, e.g., Eq.
(\ref{MLanDir})). Also, the \textquotedblleft
Schr\"{o}dinger-equivalent potential\textquotedblright\
$U_{\mathrm{SEP}}$ (Eq. (\ref{Usep})) in these models increases
linearly with nucleon energy even at high energies. Recently,
momentum-dependent nucleon self-energies have been introduced in
the RMF model by including in the Lagrangian density the couplings
of meson fields to the derivatives of nucleon densities
\cite{Typ03,Typ05}, and the results indicate that a reasonable
energy dependence of the \textquotedblleft
Schr\"{o}dinger-equivalent potential\textquotedblright\ in
symmetric nuclear matter at saturation density can be obtained,
and the Landau mass can also be increased to a more reasonable
value while keeping the Dirac mass unchanged, which further leads
to an improved description of $\beta $-decay half-lives of
neutron-rich nuclei in the $Z\approx 28$ and $Z\approx 50$ regions
\cite{Mar07}. In the framework of density-functional theory,
including the couplings of meson fields to the derivatives of
nucleon densities in the Lagrangian density provides an effective
way to take into account higher-order effects. Another way to
introduce the momentum-dependence in nucleon self-energies is to
include the Fock exchange terms by means of the relativistic
Hartree-Fock (RHF) approximation, even though in practice the
inclusion of the Fock terms would increase significantly the
numerical complexity such that it is very difficult to find
appropriate effective Lagrangians for the RHF model to give
satisfactory quantitative description of the nuclear structure
properties compared with standard RMF models
\cite{Mil74,Bro78,Jam81,Hor83,Blu87,Bou87,Lop88,Ber93,Nie01,Mar04,Lop05}.
Recently, there have been some developments in the
density-dependent RHF approach \cite{Lon06a,Lon06b,Lon07}. It is
shown that the density-dependent RHF model can describe the
properties of both finite nuclei and nuclear matter with results
comparable to those from standard RMF models. A more
phenomenological way to improve the results of RMF models is to
introduce momentum- as well as isospin-dependent form factors in
the meson-nucleon coupling constants. It has been shown in Refs.
\cite{Web92,Mar94,Sah00} that the empirically observed energy
dependence of the nuclear optical potential in symmetric nuclear
matter at saturation density can be reproduced by relativistic
mean-field models with momentum-dependent form factors. Finally,
to better understand the isospin-dependent properties of
asymmetric nuclear matter it is crucial to investigate the density
and momentum dependence of underlying isovector nuclear effective
interaction. To reach this ultimate goal, we need not only more
advanced theoretical approaches but also more experimental data
both on finite nuclei, especially those far from $\beta
$-stability line, and from heavy-ion reactions induced by high
energy neutron-rich nuclei.

\begin{acknowledgments}
We would like to thank Paolo Finelli, Wei-Zhou Jiang, Plamen G.
Krastev, and Gao-Chan Yong for helpful discussions or
communications. L.W.C. also thanks the hospitality of Texas A\&M
University at Commerce and College Station where part of the work
was done. This work was supported in part by the National Natural
Science Foundation of China under Grant Nos. 10575071 and 10675082,
MOE of China under project NCET-05-0392, Shanghai Rising-Star
Program under Grant No. 06QA14024, the SRF for ROCS, SEM of China,
the China Major State Basic Research Development Program under
Contract No. 2007CB815004 (L.W.C.), the US National Science
Foundation under Grant No. PHY-0457265, the Welch Foundation under
Grant No. A-1358 (C.M.K.), the US National Science Foundation under
Grant No. PHY-0652548, and the Research Corporation under Award No.
7123 (B.A.L.).
\end{acknowledgments}

\appendix

\section{Isospin- and momentum-dependent MDI interaction}

\label{MDI}

The isospin- and momentum-dependent MDI interaction is based on a
modified finite-range Gogny effective interaction~\cite{Das03}. In
the MDI interaction, the potential energy density $V(\rho ,\alpha
)$ of an asymmetric nuclear matter at total density $\rho $ and
isospin asymmetry $\alpha $ is expressed as
follows~\cite{Das03,Che05a},
\begin{eqnarray}
V(\rho ,\alpha ) &=&\frac{A_{u}\rho _{n}\rho _{p}}{\rho _{0}}+%
\frac{A_{l}}{2\rho _{0}}(\rho _{n}^{2}+\rho _{p}^{2})+\frac{B}{\sigma +1}%
\frac{\rho ^{\sigma +1}}{\rho _{0}^{\sigma }}  \notag \\
&\times &(1-x\alpha ^{2})+\frac{1}{\rho _{0}}\sum_{\tau ,\tau
^{\prime
}}C_{\tau ,\tau ^{\prime }}  \notag \\
&\times &\int \int d^{3}pd^{3}p^{\prime }\frac{f_{\tau }(\vec{r},\vec{p}%
)f_{\tau ^{\prime }}(\vec{r},\vec{p}^{\prime
})}{1+(\vec{p}-\vec{p}^{\prime })^{2}/\Lambda ^{2}}.  \label{MDIV}
\end{eqnarray}%
In the mean-field approximation, Eq. (\ref{MDIV}) leads to the
following single-particle potential for a nucleon with momentum
$\vec{p}$ and isospin $\tau $ in asymmetric nuclear
matter~\cite{Das03,Che05a}:
\begin{eqnarray}
U(\rho ,\alpha ,\vec{p},\tau ) &=&A_{u}(x)\frac{\rho _{-\tau }}{\rho _{0}}%
+A_{l}(x)\frac{\rho _{\tau }}{\rho _{0}}  \notag \\
&+&B(\frac{\rho }{\rho _{0}})^{\sigma }(1-x\alpha ^{2})-8\tau x\frac{B}{%
\sigma +1}\frac{\rho ^{\sigma -1}}{\rho _{0}^{\sigma }}\alpha \rho
_{-\tau }
\notag \\
&+&\frac{2C_{\tau ,\tau }}{\rho _{0}}\int d^{3}p^{\prime }\frac{f_{\tau }(%
\vec{r},\vec{p}^{\prime })}{1+(\vec{p}-\vec{p}^{\prime
})^{2}/\Lambda ^{2}}
\notag \\
&+&\frac{2C_{\tau ,-\tau }}{\rho _{0}}\int d^{3}p^{\prime }\frac{f_{-\tau }(%
\vec{r},\vec{p}^{\prime })}{1+(\vec{p}-\vec{p}^{\prime
})^{2}/\Lambda ^{2}}. \label{MDIU}
\end{eqnarray}%
In the above $\tau =1/2$ ($-1/2$) for neutrons (protons); $\sigma
=4/3$; $ f_{\tau }(\vec{r},\vec{p})$ is the phase-space
distribution function at coordinate $\vec{r}$ and momentum
$\vec{p}$. The parameters $ A_{u}(x),A_{l}(x),B,C_{\tau ,\tau
},C_{\tau ,-\tau }$ and $\Lambda $ are obtained by fitting the
momentum-dependence of $U(\rho ,\alpha ,\vec{p},\tau )$ to that
predicted by the Gogny Hartree-Fock and/or the
Brueckner-Hartree-Fock calculations, the saturation properties of
symmetric nuclear matter and the symmetry energy of $31.6$ MeV at
normal nuclear matter density $\rho _{0}=0.16$ fm$^{-3}$
\cite{Das03}. The incompressibility $K_{0}$ of cold symmetric
nuclear matter at saturation density $\rho _{0}$ is set to be
$211$ MeV. The parameters $A_{u}(x)$ and $A_{l}(x)$ depend on the
$x$ parameter according to
\begin{equation}
A_{u}(x)=-95.98-x\frac{2B}{\sigma +1},~A_{l}(x)=-120.57+x\frac{2B}{\sigma +1}%
.
\end{equation}%
The different $x$ values in the MDI interaction are introduced to
vary the density dependence of the nuclear symmetry energy while
keeping other properties of the nuclear equation of state fixed
\cite{Che05a}, and they can be adjusted to mimic the predictions
of microscopic and/or phenomenological many-body theories on the
density dependence of nuclear matter symmetry energy. The last two
terms in Eq. (\ref{MDIU}) contain the momentum-dependence of the
single-particle potential. The momentum dependence of the symmetry
potential stems from the different interaction strength parameters
$C_{\tau ,-\tau }$ and $C_{\tau ,\tau }$ for a nucleon of isospin
$\tau $ interacting, respectively, with unlike and like nucleons
in the background fields. More specifically, we use $C_{\tau
,-\tau }=-103.4$ MeV and $C_{\tau ,\tau }=-11.7$ MeV.

With $f_{\tau }(\vec{r},\vec{p})$ $=\frac{2}{h^{3}}\Theta
(p_{f}(\tau )-p)$ for nuclear matter at zero temperature, the
integrals in Eqs.~(\ref{MDIV})~and (\ref{MDIU}) can be calculated
analytically and we find
\begin{widetext}
\begin{eqnarray}
&&\int \int d^{3}pd^{3}p^{\prime }\frac{f_{\tau
}(\vec{r},\vec{p})f_{\tau ^{\prime }}(\vec{r},\vec{p}^{\prime
})}{1+(\vec{p}-\vec{p}^{\prime
})^{2}/\Lambda ^{2}}  \nonumber \\
&=&\frac{1}{6}\left( \frac{4\pi }{h^{3}}\right) ^{2}\Lambda
^{2}\left\{ p_{f}(\tau )p_{f}(\tau ^{\prime })\left[
3(p_{f}^{2}(\tau )+p_{f}^{2}(\tau
^{\prime }))-\Lambda ^{2}\right] \right.   \nonumber \\
&&+4\Lambda \left[ (p_{f}^{3}(\tau )-p_{f}^{3}(\tau ^{\prime }))\tan ^{-1}%
\frac{p_{f}(\tau )-p_{f}(\tau ^{\prime })}{\Lambda
}-(p_{f}^{3}(\tau )+p_{f}^{3}(\tau ^{\prime }))\tan
^{-1}\frac{p_{f}(\tau )+p_{f}(\tau
^{\prime })}{\Lambda }\right]   \nonumber \\
&&\left. +\frac{1}{4}\left[ \Lambda ^{4}+6\Lambda
^{2}(p_{f}^{2}(\tau )+p_{f}^{2}(\tau ^{\prime }))-3(p_{f}^{2}(\tau
)-p_{f}^{2}(\tau ^{\prime }))^{2}\right] \ln \frac{(p_{f}(\tau
)+p_{f}(\tau ^{\prime }))^{2}+\Lambda ^{2}}{(p_{f}(\tau
)-p_{f}(\tau ^{\prime }))^{2}+\Lambda ^{2}}\right\}
\end{eqnarray}
and
\begin{eqnarray}
&&\int d^{3}p^{\prime }\frac{f_{\tau }(\vec{r},\vec{p}^{\prime })}{1+(\vec{p}%
-\vec{p}^{\prime })^{2}/\Lambda ^{2}}  \nonumber \\
&=&\frac{2}{h^{3}}\pi \Lambda ^{3}\left[ \frac{p_{f}^{2}(\tau
)+\Lambda
^{2}-p^{2}}{2p\Lambda }\ln \frac{(p+p_{f}(\tau ))^{2}+\Lambda ^{2}}{%
(p-p_{f}(\tau ))^{2}+\Lambda ^{2}}\right.
+\left. \frac{2p_{f}(\tau )}{\Lambda }-2\tan ^{-1}\frac{p+p_{f}(\tau )}{%
\Lambda }-2\tan ^{-1}\frac{p-p_{f}(\tau )}{\Lambda }\right] .  \nonumber \\
&&
\end{eqnarray}
\end{widetext}
With above results as well as the well-known contribution from
nucleon kinetic energies in free Fermi gas model, we can thus
easily obtain the EOS of asymmetric nuclear matter at zero
temperature.

We note that the MDI interaction has been extensively used in the
transport model for studying isospin effects in intermediate
energy heavy-ion collisions induced by neutron-rich nuclei \cite%
{LiBA04a,Che04,Che05a,LiBA05a,LiBA05b,LiBA06b,Yon06a,Yon06b,Yon07}
and the study of the thermal properties of asymmetric nuclear
matter~\cite{Xu07,Xu07b}. In particular, the isospin diffusion
data from NSCL/MSU have constrained the value of $x$ to between
$0$ and $-1$ for nuclear matter densities less than about $
1.2\rho _{0}$ \cite{Che05a,LiBA05c}.

\end{document}